\documentclass[a4paper,11pt]{article}
\pdfoutput=1 
\usepackage{jheppub}
\usepackage{comment}
\usepackage{tikz}
\usepackage[T1]{fontenc}

\usepackage{scalerel,stackengine}
\stackMath
\newcommand\reallywidehat[1]{%
\savestack{\tmpbox}{\stretchto{%
  \scaleto{%
    \scalerel*[\widthof{\ensuremath{#1}}]{\kern-.6pt\bigwedge\kern-.6pt}%
    {\rule[-\textheight/2]{1ex}{\textheight}}
  }{\textheight}%
}{0.5ex}}%
\stackon[1pt]{#1}{\tmpbox}%
}
\usepackage{cleveref}
\newcommand{\oS}{\overline{S}}

\newcommand{\omW}{\overline{\mathcal{W}}}
\newcommand{\omJ}{\overline{\mathcal{J}}}
\newcommand{\oC}{\overline{C}}
\newcommand{\oD}{\overline{D}}
\newcommand{\oz}{\overline{z}}
\newcommand{\oT}{\overline{T}}
\newcommand{\oj}{\overline{j}}

\newcommand{\oQ}{\overline{Q}}

\newcommand{\opsi}{\overline{\psi}}

\newcommand{\otheta}{\overline{\theta}}
\newcommand{\olambda}{\overline{\lambda}}
\newcommand{\oG}{\overline{G}}
\newcommand{\oh}{\overline{h}}
\newcommand{\mJ}{\mathcal{J}}

\newcommand{\mW}{\mathcal{W}}
\newcommand{\mS}{\mathcal{S}}
\newcommand{\mT}{\mathcal{T}}
\newcommand{\mR}{\mathcal{R}}

\newcommand{\oZ}{\overline{Z}}

\newcommand{\oY}{\overline{Y}}

\usepackage{amsmath}

\title{$T \overline{T}$ deformation in SCFTs and integrable supersymmetric theories}

\author[a]{Stephen Ebert,}
\author[b,c]{Hao-Yu~Sun}
\author[d]{and Zhengdi Sun}

\affiliation[a]{Mani L. Bhaumik Institute for Theoretical Physics,\\ University of California, Los Angeles, CA 90095-1547, USA}
\affiliation[b]{Center for Theoretical Physics and Department of Physics, \\ University of California, Berkeley, CA 94720-7300, USA}
\affiliation[c]{Department of Physics, University of Texas, Austin, TX 78712-1192, USA}
\affiliation[d]{Department of Physics, University of California, San Diego, CA 92093-0319, USA}

\emailAdd{stephenebert@physics.ucla.edu}
\emailAdd{hkdavidsun@utexas.edu}
\emailAdd{z5sun@ucsd.edu}

\abstract{We calculate the $\mS$-multiplets for two-dimensional Euclidean $\mathcal{N}=(0,2)$ and $\mathcal{N} = (2,2)$ superconformal field theories under the $T\oT$ deformation at leading order of perturbation theory in the deformation coupling. Then, from these $\mathcal{N} = (0, 2)$ deformed multiplets, we calculate two- and three-point correlators. We show the $\mathcal{N} = (0,2)$ chiral ring's elements do not flow under the $T\oT$ deformation. Specializing to integrable supersymmetric seed theories, such as $\mathcal{N} = (2,2)$ Landau-Ginzburg models, we use the thermodynamic Bethe ansatz to study the S-matrices and ground state energies. From both an S-matrix perspective and Melzer's folding prescription, we show that the deformed ground state energy obeys the inviscid Burgers’ equation. Finally, we show that several indices independent of $D$-term perturbations including the Witten index, Cecotti-Fendley-Intriligator-Vafa index and elliptic genus do not flow under the $T\oT$ deformation.}

\begin{document} 
\maketitle
\flushbottom
\newpage
\section{Introduction}
\label{sec:intro}
\subsection{Background}
Recent attention has been drawn toward irrelevant deformations of two-dimensional quantum field theories (QFTs) and applications in holography. Unlike marginal or relevant deformations, irrelevant deformations in QFTs are notoriously arduous to study due to the requirement of including infinitely many counterterms to the action, and understanding the ultraviolet physics of the system becomes highly ambiguous. It was not until Zamolodchikov shed light on this subject matter by deriving his novel composite operator $T\oT = \det T_{\mu \nu}(\lambda)$ \cite{Zamolodchikov:2004ce} which circumvents this counterterm technicality that the deformed Euclidean two-dimensional QFT is solvable as a function of the deformation coupling $\lambda$. The $T\oT$ deformation is a double trace operator\footnote{There is also a single trace version of the $T\oT$ deformation with applications to string theory in the bulk interpolating between AdS$_3$ in the IR and a linear dilaton spacetime in the UV \cite{Giveon:2017nie, Giveon:2017myj, Chakraborty:2020cgo}.} defined by solving the following ordinary differential equation for the deformed action $S(\lambda)$

\begin{equation}
\label{eq:definitionofTT}
   \frac{dS}{d \lambda} = - \int d^2x \sqrt{g} T\oT (x),
\end{equation}
where the deformed stress tensor is $T_{\mu \nu}[S(\lambda)]$. The $T\oT$ operator on a cylinder or flat plane is
\begin{equation}
    T\oT :=\lim _{z^{\prime} \rightarrow z}\left[T\left(z^{\prime}\right) \bar{T}(z)-\Theta\left(z^{\prime}\right) \Theta(z)\right],
\end{equation}
where in complex coordinates $(z,\bar{z})$, we have defined the standard conventions: 
\begin{equation}
    T= T_{zz},\quad \oT = T_{\bar{z} \bar{z}},\quad  \Theta = -T_{z\bar{z}}.
\end{equation}

Using conservation of the stress tensor $\nabla^\mu T_{\mu \nu} = 0$ and assuming the undeformed theory to be a CFT, we arrive at the following trace flow equation
\begin{equation}
\label{eq:traceflow}
    T^i_i = -  \pi \lambda T\oT + \cdots,
\end{equation}
where $\cdots$ are total derivatives of local operators $\nabla_z O_\alpha (z)$. Notice that these additional total derivative terms arise at least $O(\lambda^2)$ because the operator product defining the $T\overline{T}$ deformation is non-singular when the seed theory is a CFT. We will take advantage of \eqref{eq:traceflow} being exact at the leading order of $\lambda$ since we are working with the conformal perturbative theory near the SCFT fixed point. 

More recently, the $T\oT$ deformation was studied in the context of the S-matrix and finite volume spectrum \cite{Smirnov:2016lqw, Cavaglia:2016oda}. One of the many novelties the authors in \cite{Smirnov:2016lqw, Cavaglia:2016oda} found was that the $T\oT$ deformation showcases non-local and string-theoretic properties. For instance, the non-locality can be seen through the most popular example by deforming a seed action of $N$ free scalars. Using (\ref{eq:definitionofTT}), one finds the deformed action is the Nambu-Goto action in the static gauge with manifest $SO(N+2)$ symmetry. Meanwhile, one can also show that the energy of a generic state in the deformed theory is characterized by the inviscid Burgers' equation. \footnote{One also could generalize this argument by coupling the seed scalar theory to an arbitrary background metric as \cite{Bonelli:2018kik} did and show the deformed energy spectrum still obeys the inviscid Burgers' equation.} By placing the deformed theory on a torus, modular invariance would further imply the density of states exhibit a Hagedorn behavior, i.e., $\rho(E) \sim e^{-E}$, which also hints at non-locality. There are other examples one could see this non-locality property. Another notable example is that when one $T\oT$ deforms the seed action of a Yang-Mills gauge field coupled to a scalar, the deformed theory gives a non-abelian analogue of a two-dimensional DBI action \cite{Brennan:2019azg}.

As a matter of fact, we will see the same inviscid Burgers' equation in the deformed free scalar field theory used in \S \ref{sec:S-matrix} when we calculate the flow equation of the ground state energy for deformed integrable supersymmetric field theories and common supersymmetric indices. 

Shortly after findings in \cite{Smirnov:2016lqw, Cavaglia:2016oda}, the $T\oT$ deformation saw applications in the AdS$_3$/CFT$_2$ correspondence by McGough, Mezei and Verlinde \cite{McGough:2016lol}. The $T\oT$ deformation puts the bulk at a finite cutoff whose position is controlled by the deformation coupling $\lambda$. The deformation clearly spoils conformal invariance of the two-dimensional boundary field theory, but, surprisingly, remains holographically dual to the three-dimensional bulk. Further evidence is supported by Kraus, Liu and Marolf \cite{Kraus:2018xrn} where, at $O(\lambda^2)$ and $O(\lambda)$, they perfectly matched the two- and three-point bulk and boundary deformed stress tensor correlators respectively. More recently, understanding the $T\oT$ deformation in curved spacetime for AdS$_2$ \cite{Jiang:2019tcq, Brennan:2020dkw} and AdS$_3$ \cite{Mazenc:2019cfg, Caputa:2020lpa, Kraus:2021cwf} has greatly improved.

There are an abundance of applications of the $T\oT$ deformation\footnote{We refer the reader to this set of lecture notes \cite{Jiang:2019hxb} for a comprehensive review and several applications on the $T\oT$ deformation.}, but for the main scope of this paper, we focus on two-dimensional Euclidean $\mathcal{N} = (0,2)$ and $\mathcal{N} = (2,2)$ SCFTs as well as some $\mathcal{N} = (2,2)$ integrable theories. The $T\oT$ deformation for manifest supersymmetric two-dimensional QFTs were first initiated by extending (\ref{eq:definitionofTT}) to be constructed as a supersymmetric descendant $O(\zeta)$ from the supercurrent multiplet. For example, in $\mathcal{N} = (1,1)$ supersymmetry, let $(\mathcal{J}_{+++}, \mathcal{J}_{-})$ and $(\mathcal{J}_{---}, \mathcal{J}_+)$ belong to a supercurrent multiplet so the deformed superspace action is \cite{Baggio:2018rpv, Chang:2018dge}
\begin{equation}
   S(\lambda) = S(0) + \lambda \int d^2z \int d\theta^+ d\theta^- O (\zeta),
\end{equation}
where $O(\zeta) = \mathcal{J}_{+++}(\zeta) \mathcal{J}_{---}(\zeta)-\mathcal{J}_{-}(\zeta) \mathcal{J}_{+}(\zeta)$ is well-defined up to equations of motion and total derivative terms. Therefore, since the $T\oT$ deformation can be built out of a supersymmetric descendant from an $\mathcal{N} = (1,1)$ supermultiplet, it preserves all of the supersymmetry as well as integrability.

The $T\oT$ deformation was shown to be solvable and preserve the original $\mathcal{N} = (0,1)$ and $\mathcal{N} = (1,1)$ \cite{Baggio:2018rpv, Chang:2018dge, Coleman:2019dvf}, $\mathcal{N} = (0,2)$ \cite{Jiang:2019hux} and $\mathcal{N} = (2,2)$ \cite{Chang:2019kiu, Ferko:2019oyv} supersymmetries. Moreover, there are recent studies of other irrelevant deformations, such as $J\overline{T}$ or $\bar{J} T$ \cite{Guica:2017lia, Aharony:2018bad, Guica:2019vnb, LeFloch:2019rut, He:2019vzf, Aguilera-Damia:2019tpe, Guica:2020uhm} that break Lorentz invariance, in the context of supersymmetry \cite{Jiang:2019trm}.

From these recent developments on the $T\oT$ deformation in supersymmetric field theories, one can in principle calculate correlators associated to each $\mathcal{N} = (p,q)$ supersymmetric theory via superconformal Ward identities. For example, the authors in \cite{He:2019ahx} perturbatively calculated the $n$-point correlators of two-dimensional $\mathcal{N} = (1,1)$ and $\mathcal{N} = (2,2)$ superconformal field theories (SCFTs).

\subsection{Main results and organization}
\label{sec:outline}
The main results and organization of this paper are the following.

In \S \ref{sec:(0,2)}, we first embed the trace flow equation (\ref{eq:traceflow}) into the $\mathcal{R}$- and $\mathcal{S}$-multiplets for a two-dimensional $\mathcal{N} = (0,2)$ SCFT. Equipped with these deformed multiplets, we calculate the deformed two- and three-point correlators using methods developed in \cite{Dumitrescu:2011iu, Kraus:2018xrn, He:2019ahx}.  We perform conformal perturbation theory to derive deformed $n$-point correlators for $n$ general supermultiplets at $O(\lambda)$ following \cite{Cardy:2019qao} and show that the two-dimensional $\mathcal{N} = (0,2)$ SCFT's chiral ring elements do not flow under the deformation. 

In \S \ref{sec:(2,2)}, we repeat the same analysis for the deformed $\mS$-multiplet as in \S \ref{sec:(0,2)} but for $\mathcal{N} = (2,2)$ superconformal symmetry and conveniently list all of the deformed $\mS$-multiplet's components in appendix \ref{App:22}. We find that naively the $T\oT$ deformation breaks both $U(1)_A$ and $U(1)_V$ R-symmetries, so the $\mS$-multiplet becomes the generic one without superconformal symmetry. However, there exist improvement transformations which allows one to restore the conservation of one of the R-symmetries. Hence the $\mS$-multiplets can either be improved to the Ferrara-Zumino (FZ)-multiplet or $\mR$-multiplet. This is consistent with the usual expectation for a non-conformal supersymmetric theory. We additionally find one of the central currents $Y_{\pm\pm}$ or $G_{\pm\pm}$ will be generated. The central currents take the form of a total derivative; however, it could still lead to non-trivial charge for non-perturbative configurations. This is analogous to the instanton number in four-dimensional gauge theory: $\frac{1}{8\pi^2} \int \operatorname{Tr} F\wedge F$. We take this as a hint that one must further study the non-perturbative effects of the $T\oT$ deformation to completely understand the perturbative structure of the $\mS$-multiplet. One might find it tempting to conclude the chiral ring or twisted chiral ring will cease to exist in the deformed theory based on the generation of the central current. However, it is ambiguous whether every would-be chiral or twisted chiral ring elements would actually be charged under the central current. We believe understanding the non-perturbative effect of the $T\oT$ deformation and perhaps a model dependent analysis are required to determine the ultimate fate of the chiral ring and twisted chiral ring in the deformed theory.

In \S \ref{sec:S-matrix}, we find the deformed S-matrix and ground state energy for two-dimensional $\mathcal{N} = (2,2)$ Landau-Ginzburg models with superpotential $W (X, \beta) = \frac{X^{n+1}}{n+1} - \beta X$ using the thermodynamic Bethe ansatz (TBA). We show that the deformed ground state energy obeys the inviscid Burgers’ equation, and perturbatively calculate the ground state energy for each soliton system to leading order in the radius of the spatial circle. One can also generalize this analysis to some $\mathcal{N} = (1,1)$ theory using Melzer's folding prescription \cite{Melzer:1994qp}, which relates the TBA analyses for integrable $\mathcal{N} = (2,2)$ models to the corresponding $\mathcal{N} = (1, 1)$ ones. This allows us to show the deformed ground state energy of $\mathcal{N} = (1,1)$ integrable models obey the inviscid Burgers' equation and confirm the folded integrable $\mathcal{N} = (2,2)$ models' deformed ground state energy matches exactly with the deformed ground state energy in $\mathcal{N} = (1,1)$ integrable models.  

In the same section, we also explore well-studied supersymmetric indices under the $T\overline{T}$ deformation and derive their corresponding flow equations via TBA. In particular, we show the Witten index, Cecotti-Fendley-Intriligator-Vafa (CFIV) index and elliptic genus do not flow under the $T\oT$ deformation in the integrable supersymmetric theories. This is perhaps not a surprise because in general these quantities do not depend on the $D$-term deformation. More generally, we expect quantities dependent on $D$-terms like $\text{Tr} (-1)^F F^l e^{-\beta H}$ for $l > 1$ to flow under the deformation. This is consistent with the results by \cite{Chang:2019kiu}, where they showed the K\"{a}hler potential ($D$-term) receives corrections from the $T\oT$ deformation while the superpotential ($F$-term) is protected. However, it is also worth to keep in mind that for a generic supersymmetric field theory, non-perturbative effects of the $T\oT$ deformation may lead to some ``large'' $D$-term deformation which will change the index. Also, the CFIV index not flowing hints at the possibility that the $\mathcal{N} = (2, 2)$ chiral ring do not flow. But we will leave a thorough study on these two questions and other directions for future works.

To conclude this paper, in \S \ref{sec:conclusion} we will discuss open questions for future directions including: studying SCFT correlators under different irrelevant deformations, a possible way to derive the $tt^*$ equations using the deformed CFIV index and TBA analyses for other deformed supersymmetric integrable models, and reflection matrices in the presence of a boundary.

\section{Deforming two-dimensional \texorpdfstring{$\mathcal{N} = (0,2)$}{} SCFT}
\label{sec:(0,2)}
In this section, we extend the analysis by Kraus, Liu and Marolf \cite{Kraus:2018xrn} to the two-dimensional $\mathcal{N} = (0,2)$ SCFT setting. A main feature of a supersymmetric field theory is that various bosonic and fermionic operators can combine into supermultiplets, which are representations of the supersymmetry algebra. In particular, the stress tensor $T_{\mu\nu}$ is embedded into the $\mS$-multiplet introduced by Dumitrescu and Seiberg \cite{Dumitrescu:2011iu} and may be reduced to a smaller multiplet such as the $\mR$-multiplet. We will review this structure, and then at leading order in $\lambda$, derive the two-point correlator of such supermultiplet using perturbation theory. Besides the constraints from stress tensor conservation and rotational/translational invariance used in \cite{Kraus:2018xrn}, we also need to exploit constraints from $\mathcal{N} = (0,2)$ supersymmetry. As we see, this is easily achieved by embedding the operator equation (\ref{eq:traceflow})
\begin{equation}
\label{eq:flow}
    T_{z\oz} = - \pi \lambda T_{zz}T_{\oz\oz} + O(\lambda^2)
\end{equation}
into an operator equation written in superspace
\begin{equation}
\label{eq:superspaceTTbar}
    \mathcal{R}_{--} = \frac{\pi\lambda}{16} \mathcal{R}_{++} \mT_{----} + O(\lambda^2).
\end{equation}
Then, by taking each component of (\ref{eq:superspaceTTbar}), we derive an operator equation that allows us to calculate two- and three-point correlators as \cite{Kraus:2018xrn} did at leading order in $\lambda$. 

\subsection{Deformed stress tensor correlators in non-supersymmetric CFT}
\label{ReviewKLM}
To be self-contained, we go over the pivotal results in \cite{Kraus:2018xrn} for the 2- and 3-point stress tensor correlators in a non-supersymmetric $T\overline{T}$-deformed CFT. As previously mentioned, we make use of rotational/translational symmetries to constrain the form of the two-point functions
\begin{equation}
    \begin{aligned}
&\left\langle T_{z z}(z) T_{z z}(0)\right\rangle_{\lambda}=\frac{f_{1}(y)}{z^{4}},  \\
&\left\langle T_{z z}(z) T_{z \bar{z}}(0)\right\rangle_{\lambda}=\frac{f_{2}(y)}{z^{3} \bar{z}},  \\
&\left\langle T_{z z}(z) T_{\overline{z z}}(0)\right\rangle_{\lambda}=\frac{f_{3}(y)}{z^{2} \bar{z}^{2}}, \\
&\left\langle T_{z \bar{z}}(z) T_{z \bar{z}}(0)\right\rangle_{\lambda}=\frac{f_{4}(y)}{z^{2} \bar{z}^{2}} ,
\end{aligned}
\end{equation}
where  $\left \langle \cdots \right \rangle_\lambda$ corresponds to the deformed correlator and  $y = \frac{|z|^2}{\lambda}$ is a dimensionless coordinate. We can solve for these unknown functions via stress tensor conservation to find three ordinary differential equations:
\begin{equation}
    \begin{aligned}
    \label{diffeqn}
&f_{1}^{\prime}+y^{3}\left(\frac{f_{2}}{y^{3}}\right)^{\prime}=0, \\
&\left(\frac{f_{2}}{y}\right)^{\prime}+y\left(\frac{f_{3}}{y^{2}}\right)^{\prime}=0, \\
&\left(\frac{f_{2}}{y}\right)^{\prime}+y\left(\frac{f_{4}}{y^{2}}\right)^{\prime}=0,
\end{aligned}
\end{equation}
where $\prime = \frac{d}{dy}$ and the initial conditions are for a CFT:
\begin{equation}
    f_{1} \rightarrow \frac{c}{2}, \quad \frac{1}{y^{2}} f_{2} \rightarrow 0, \quad \frac{1}{y^{2}} f_{3} \rightarrow 0, \quad \frac{1}{y^{2}} f_{4} \rightarrow 0, \quad y \rightarrow \infty.
\end{equation}
However, stress tensor conservation is insufficient to uniquely fix all four unknown functions. We can use conformal perturbation theory to determine one of the functions and then substitute that result into the differential equations coming from stress tensor conservation to determine the other three unknown functions. An important observation is to note that all two-point correlators involving the stress tensor start to receive corrections at $O(\lambda^2)$. At $O(\lambda^2)$, using conformal perturbation theory and \eqref{eq:flow}, $ \left \langle T_{z\bar{z}}(z) T_{z \bar{z}} (0) \right \rangle_\lambda$ is:
\begin{equation}
    \begin{aligned}
    \label{O(lambdasq)}
         \langle T_{z\bar{z}}(z) T_{z \bar{z}} (0) \rangle_\lambda &=  \frac{\pi^2 \lambda^2 c^2}{4} \frac{1}{z^4 \bar{z}^4} \implies f_4(y) = \frac{\pi^2 c^2}{4y^2}.
    \end{aligned}
\end{equation}
Substituting \eqref{O(lambdasq)} into \eqref{diffeqn}, we automatically determine the other 3 unknown functions. Putting everything together, we arrive at
\begin{equation}
    \begin{aligned}
\left\langle T_{z z}(z) T_{z z}(0)\right\rangle_{\lambda} &=\frac{c}{2 z^{4}}+\frac{5 \pi^{2} \lambda^{2} c^{2}}{6} \frac{1}{z^{6} \bar{z}^{2}}+O(\lambda^3), \\
\left\langle T_{z z}(z) T_{z \bar{z}}(0)\right\rangle_{\lambda} &=-\frac{\pi^{2} \lambda^{2} c^{2}}{3} \frac{1}{z^{5} \bar{z}^{3}}+O(\lambda^3), \\
\left\langle T_{z z}(z) T_{\overline{z z}}(0)\right\rangle_{\lambda} &=\frac{\pi^{2} \lambda^{2} c^{2}}{4} \frac{1}{z^{4} \bar{z}^{4}}+O(\lambda^3), \\
\left\langle T_{z\bar{z}} (z) T_{z \bar{z}}(0)\right\rangle_{\lambda} &=\frac{\pi^{2} \lambda^{2} c^{2}}{4} \frac{1}{z^{4} \bar{z}^{4}}+O(\lambda^3).
\end{aligned}
\end{equation}
Likewise, one can also use conformal perturbation theory along with \eqref{eq:flow} to calculate three-point correlators and find 
\begin{equation}
\begin{aligned}
    &\left\langle T_{z \bar{z}}\left(z_{1}\right) T_{z z}\left(z_{2}\right) T_{\bar{z} \bar{ z}}\left(z_{3}\right)\right\rangle_{\lambda}=-\frac{\pi \lambda c^{2}}{4} \frac{1}{\left(z_{1}-z_{2}\right)^{4}\left(\bar{z}_{1}-\bar{z}_{3}\right)^{4}}+O\left(\lambda^{2}\right),  \\
   &\left \langle T_{zz} (z_1) T_{\bar{z} \bar{z}}(z_2) T_{\bar{z} \bar{z}}(z_3) \right \rangle_\lambda = -\frac{\pi \lambda c^{2}}{3} \frac{1}{\left(z_{1}-z_{2}\right)^{3}} \frac{1}{\left(\bar{z}_{2}-\bar{z}_{3}\right)^{5}} - \frac{\pi \lambda c^2}{3} \frac{1}{(z_{3} -z_1)^3 (\bar{z}_2 - \bar{z}_3)^5}  + O(\lambda^2), \\ 
   &\left \langle T_{zz} (z_1) T_{zz}(z_2) T_{zz}(z_3) \right \rangle_ \lambda = \frac{c}{(z_{1} - z_{2})^2 (z_{2} - z_{3})^2 (z_{3} - z_{1})^2} + O(\lambda^2).
   \end{aligned}
\end{equation}

In the subsequent sections, we perform a similar analysis as presented in this subsection for the supersymmetric cases involving their $\mathcal{R}$- and $\mathcal{S}$-multiplets.

\subsection{The deformed \texorpdfstring{$\mS$}{}-multiplet's structure}
In this subsection, we present a short review on the $\mS$-multiplet in Euclidean signature\footnote{Note that notations in \cite{Dumitrescu:2011iu} are in Lorentzian signature.} together with an analysis on the $\mS$-multiplet's structure under the $T\oT$ deformation. 

We define the supercovariant derivatives as 
\begin{equation}
    D = \partial_\theta + \otheta \partial_z, \quad \oD = \partial_{\otheta} + \theta \partial_z.
\end{equation}

For a general $\mathcal{N} = (0,2)$ supersymmetric theory, its $\mS$-multiplet is \cite{Dumitrescu:2011iu}: 

\begin{equation}
\begin{aligned}
\label{S-multiplet(0,2)}
    \mS_{z} &= j_z + i \otheta S_{z} + i \theta \oS_{z} + 2 \theta \otheta T_{zz}, \\
    \mW_{\oz} &= - \frac{i}{2}S_{\oz} + \theta \left(T_{z\oz} + \frac{1}{2}\partial_{\oz} j_{z}\right) + \otheta C - \frac{i}{2}\theta \otheta \partial_z S_{\oz}, \\
    \omW_{\oz} &= \frac{i}{2}\oS_{\oz} + \theta \oC + \otheta\left(T_{z\oz} - \frac{1}{2}\partial_{\oz} j_{z}\right) - \frac{i}{2}\theta \otheta \partial_z\oS_{\oz}, \\
    \mT_{\oz\oz} &= T_{\oz\oz} - \frac{i}{2}\theta \partial_{\oz}\oS_{\oz} + \frac{i}{2}\otheta \partial_{\oz}S_{\oz} + \frac{1}{2}\theta\otheta \partial_{\oz}^2 j_z,
\end{aligned}
\end{equation}
where the components of the conserved stress tensor are $T_{zz},\,T_{z\oz},\,T_{\oz\oz}$, the supersymmetric current are\footnote{All of the conserved supersymmetric currents' components have the same chirality, so we drop the spinor indices for convenience.} $S_{z},\,S_{\oz},\,\oS_{z},\,\oS_{\oz}$, the $R$-current being an integral of the sum of $j_z$ and $j_{\bar{z}}$, and the complex constant $C$ is to be interpreted as a space-filling brane current.

The general solution for the defining equations of the $\mS$-multiplet \eqref{S-multiplet(0,2)}
\begin{equation}
\begin{aligned}
    \partial_{\oz} \mS_{z} &= D \mW - \oD \omW, \\
    \oD \mW &= C, \,\,\, D \omW = \oC, \\
    \oD \mT_{\oz\oz} &=  - \partial_{\oz} \mW_{\oz}, \,\,\, D \mT_{\oz\oz} = - \partial_{\oz} \omW_{\oz}.
\end{aligned}
\end{equation}
 For $\mathcal{N} = (0,2)$ supersymmetry, the supercurrent multiplet contains two real supercurrents $S_1,S_2$ with dimension $3/2$. We write $S_1, S_2$ as a complex supercurrent by defining $S_\mu = (S_1)_\mu + i (S_2)_\mu$ and $\oS_\mu = (S_1)_\mu - i (S_2)_\mu$. This could potentially lead to confusion as $\oS_z \neq S_{\oz} \neq \oS_{\oz}$ and to avoid this confusion, we include $z,\oz$ indices throughout this paper.

The conservation equations are given by
\begin{equation}
\begin{aligned}
     \partial_{z}S_{\oz} + \partial_{\oz} S_{z} &= 0, \\
     \partial_{z}\oS_{\oz} + \partial_{\oz} \oS_{z} &= 0, \\
     \partial_{z}T_{\oz \mu} + \partial_{\oz} T_{z \mu} &= 0. 
\end{aligned}
\end{equation}

Next, we discuss the cases where the $\mS$-multiplet can reduce to some simple multiplet. Firstly, if $C=0$ and there exists a well-defined $j_{\oz}$ such that $\partial_{z}j_{\oz}+\partial_{\oz}j_{z} = 0$, then the $\mS$-multiplet can be improved to the $\mR$-multiplet:
\begin{equation}
\begin{aligned}
    \mR_{z} &= j_{z} + i \theta \oS_{z} + i \otheta S_{z} + 2 \theta \otheta T_{zz}, \\
    \mR_{\oz} &= j_{\oz} + i \theta \oS_{\oz} + i \otheta S_{\oz} + 2 \theta \otheta T_{z\oz}, \\
    \mT_{\oz\oz} &= T_{\oz\oz} - \frac{i}{2}\theta \partial_{\oz} \oS_{\oz} + \frac{i}{2} \otheta \partial_{\oz} S_{\oz} + \frac{1}{2} \theta \otheta \partial_{\oz}^2 j_{z},
\end{aligned}
\end{equation}
and one can check the $\mR$-multiplet satisfies the following constraints
\begin{equation}
\begin{aligned}
    \partial_{z} \mR_{\oz} + \partial_{\oz} \mR_{z} &= 0, \\
    D\left(T_{\oz\oz} + \frac{1}{2}\partial_{\oz}\mR_{\oz}\right) &= 0, \\
    \oD\left(T_{\oz\oz} - \frac{1}{2}\partial_{\oz}\mR_{\oz}\right) &= 0.
\end{aligned}
\end{equation}

Secondly, if the seed IR theory is superconformal, then all the currents are holomorphic (e.g., $j_{\oz} = S_{\oz} = \oS_{\oz} = T_{z\oz} = 0$). As a result, at the superconformal point, $\mR_{\oz}$ vanishes and the $\mS$-multiplet reduces to the holomorphic supercurrent
\begin{equation}
    \partial_{\oz}\mS_{z} = 0
\end{equation}
and the anti-holomorphic component $T_{\oz\oz}$ of the stress tensor. 

The non-trivial OPE can be nicely packaged using superspace notations \cite{Benini:2013cda}:
\begin{equation}
\label{OPEsuperspace}
\begin{aligned}
    S_z(Z_1)S_z(Z_2) & \sim \frac{\theta_{12}}{Z_{12}} DS_z(Z_2) - \frac{\otheta_{12}}{Z_{12}} \oD S_z(Z_2) + 2 \frac{\theta_{12}\otheta_{12}}{Z_{12}^2} S_z(Z_2) + 2 \frac{\theta_{12}\otheta_{12}}{Z_{12}} \partial_z S_z(Z_2) + \frac{c}{3Z_{12}^2}, \\
    T_{\oz\oz}(z_1) T_{\oz\oz}(z_2) & \sim \frac{c}{2\oz^4} + \frac{2T_{\oz\oz}(0)}{\oz^2} + \frac{\partial_{\oz} T_{\oz\oz}(0)}{\oz},
\end{aligned}
\end{equation}
where $\theta_{12} = \theta_1 - \theta_2$, $\otheta_{12} = \otheta_1 - \otheta_2$ and $Z_{12} = z_{12} - \theta_1\otheta_2 - \otheta_1\theta_1$. 

In components, we can alternatively write (\ref{OPEsuperspace}) as
\begin{equation}
\label{OPEcomponents}
\begin{split}
    T_{zz}(z)T_{zz}(0)  &\sim \frac{c}{2z^4} + \frac{2T_{zz}(0)}{z^2} + \frac{\partial_z T_{zz}(0)}{z},\\
    T_{\oz\oz}(z) T_{\oz\oz}(0) &\sim \frac{c}{2\oz^4} + \frac{2T_{\oz\oz}(0)}{\oz^2} + \frac{\partial_{\oz} T_{\oz\oz}(0)}{\oz},\\
    T_{zz}(z)j_z (0) &\sim \frac{j_z(0)}{z^2} + \frac{\partial_z j_z(0)}{z}, \\
    S_z(z)\oS_z(0) & \sim \frac{2c}{3z^3} + \frac{2j_z(0)}{z^2} + \frac{2T_{zz}(0) + \partial_z j_z(0)}{z},\\
    S_z(z) S_z(0)  & \sim \oS_z(z)\oS_z(z) \sim 0,
\end{split}
\quad
\begin{split}
    T_{zz}(z)S_z(0)  &\sim \frac{3S_z(0)}{2z^2} + \frac{\partial_z S_z(0)}{z},\\
     T_{zz}(z) \oS_z(0) &\sim \frac{3\oS_z(0)}{2z^2} + \frac{\partial_z \oS_z(0)}{z},\\
     j_z(z) S_z(0) & \sim \frac{S_z(0)}{z},\\
     j_z(z) \oS_z(0) &\sim \frac{\oS_z(0)}{z},\\
     j_z(z) j_z(0)  &\sim \frac{c}{3z^2}.
\end{split}
\end{equation}

Now, say we start with a seed supersymmetric theory which flows to a superconformal point whose superconformal $R$-symmetry is not an accidental symmetry in the IR. Then we parameterize the end of the RG flow using a parameter, $\lambda$, such that at $\lambda=0$ we are at the superconformal point. As we claimed before, all the anti-holomorphic components of the conserved currents $T,S,\oS,j$ vanish, so perturbatively in $\lambda$ they admit an expansion in terms of the SCFT's operators. For instance, under the $T\oT$ deformation, we have (\ref{eq:flow}).

In the case of a $\mathcal{N} = (0,2)$ theory, we can construct the $\mR$-multiplet perturbatively in $\lambda$ via
\begin{equation}
\label{eq:RMult}
    \mR_{\oz} = - \pi\lambda \mS^{(0)}_{z} T_{\oz\oz}^{(0)} + O(\lambda^2),
\end{equation}
where by the superscript $(0)$, we mean the $\mS_{z}$ and $T_{\oz\oz}$ in the seed SCFT. The top component of (\ref{eq:RMult}) yields the expansion of $T_{z\oz}$ in the non-supersymmetric $T\oT$ trace flow equation \eqref{eq:traceflow}. Hence we conclude, at least perturbatively in the $T\oT$ deformed theory, a generic $\mS$-multiplet reduces to the $\mR$-multiplet. The above structure agrees with deformation of the conserved currents which Cardy discovered \cite{Cardy:2019qao}.

Notice we can have a case where the irrelevant deformation can generate non-perturbative effects which explicitly breaks some of the $U(1)$ symmetries.\footnote{We thank Ken Intriligator for pointing this example out for us.} For example, if the IR theory contains two $U(1)$ symmetries with a mixed `t Hooft anomaly, we can consider gauging one of them. The $U(1)$ gauge coupling is IR free, so it is an irrelevant deformation from the point of view of the original theory. Due to the mixed `t Hooft anomaly, instanton effects of the $U(1)$ gauge field will break the other $U(1)$ symmetry non-perturbatively as a special case of UV/IR mixing! Perhaps it is too hasty to conclude that the $T\oT$ deformation will preserve the $U(1)_R$ symmetry even non-perturbatively. However, since our calculations are perturbative in $\lambda$, we will assume that the $\mS$-multiplet reduces to the $\mR$-multiplet and leave the question whether the $U(1)_R$ is broken by the $T\oT$ deformation non-perturbatively or not for future investigations.

\subsection{Deformed two-point correlators}
We extract the SCFT two-point functions from the OPEs (\ref{OPEsuperspace}) and (\ref{OPEcomponents}):
\begin{equation}
\begin{aligned}
\label{SCFTOPEs}
    \langle \mS_z(Z_1) \mS_z(Z_2) \rangle_0 = \frac{c}{3Z_{12}^2}.
\end{aligned}
\end{equation}
In components, (\ref{SCFTOPEs}) is
\begin{equation}
\begin{aligned}
    \langle T_{zz}(z) T_{zz}(0) \rangle_0 &= \frac{c}{2z^4}, \\ \langle T_{\oz\oz}(z) T_{\oz\oz}(0) \rangle_0 &= \frac{c}{2\oz^4}, \\ \langle S_z(z) \oS_z(0) \rangle_0 &= \frac{2c}{3z^3}, \\ \langle j_z(z)j_z(0) \rangle_0 &= \frac{c}{3z^2}.
\end{aligned}
\end{equation}

To derive the deformed two-point functions, we follow the method in \cite{Kraus:2018xrn} and \S \ref{ReviewKLM}. Since we are working in perturbation theory, all new terms appearing in the correlation functions must be a function of $y = \frac{z\oz}{\lambda}$ and vanish in the limit $\lambda \rightarrow 0$. Also, all the correlation functions we consider are at separate spacetime points, so we will drop present contact terms. The operator equations
\begin{equation}
    \mR_{\oz} = - \pi\lambda \mS^{(0)}_{z} T_{\oz\oz}^{(0)} + O(\lambda^2)
\end{equation}
allow us to derive the deformed two-point functions in superspace formalism
\begin{equation}
\begin{aligned}
    \langle \mR_{\oz}(Z_1) \mR_{\oz}(Z_2) \rangle_\lambda &= \pi^2 \lambda^2 \langle \mR_{z}(Z_1) \mR_{z}(Z_2) \rangle_0 \langle T_{\oz\oz}(z_1) T_{\oz\oz}(z_2) \rangle_0 + O(\lambda^3) \\& = \frac{\pi^2\lambda^2c^2}{6} \frac{1}{Z_{12}^2\oz_{12}^4} + O(\lambda^3).
    \end{aligned}
\end{equation}
Then, from the conservation equation $\partial_z \mR_{\oz} + \partial_{\oz} \mR_{z} = 0$, we find
\begin{equation}
\begin{aligned}
\label{otherRviaconservation}
    \langle \mR_{z}(Z_1) \mR_{\oz}(Z_2) \rangle_\lambda &= - \frac{\pi^2\lambda^2c^2}{9}\frac{1}{Z_{12}^3\oz_{12}^3} + O(\lambda^3), \\
    \langle \mR_{z}(Z_1) \mR_{z}(Z_2) \rangle_\lambda &= \frac{c}{3Z_{12}^2} - \frac{\pi^2\lambda^2c^2}{6} \frac{1}{Z_{12}^4\oz_{12}^2} + O(\lambda^3).
\end{aligned}
\end{equation}

We collect the non-zero two-point functions in component fields below:
\begin{equation}
\begin{aligned}
	\langle j_{\oz} (z) j_{\oz}(0) \rangle_\lambda &= \frac{\pi^2\lambda^2 c^2}{6}\frac{1}{z^2\oz^4} + O(\lambda^3), \\
    \langle j_z(z) j_{\oz}(0) \rangle_\lambda &= -\frac{\pi^2\lambda^2c^2}{9} \frac{1}{z^3\oz^3} + O(\lambda^3), \\
    \langle j_z(z) j_z(0) \rangle_\lambda &= \frac{c}{3z^2} + \frac{\pi^2\lambda^2 c^2}{6}\frac{1}{z^4\oz^2} + O(\lambda^3),
\end{aligned}
\end{equation}
\begin{equation}
\label{GCorrs}
\begin{aligned}
	\langle S_{\oz}(z) \oS_{\oz}(0) \rangle_\lambda &= \frac{\pi^2 c^2\lambda^2}{3} \frac{1}{z^3\oz^4} + O(\lambda^3), \\
    \langle S_{z}(z) \oS_{\oz}(0) \rangle_\lambda &= -\frac{\pi^2 c^2 \lambda^2}{3} \frac{1}{z^4\oz^3}+ O(\lambda^3), \\
    \langle S_{\oz}(z) \oS_{z}(0) \rangle_\lambda &= - \frac{\pi^2 c^2 \lambda^2}{3} \frac{1}{z^4 \oz^3}+ O(\lambda^3), \\
    \langle S_{z}(z) \oS_{z}(0) \rangle_\lambda &= \frac{2c}{3z^3} + \frac{2\pi^2 \lambda^2 c^2}{3} \frac{1}{z^5\oz^2}+ O(\lambda^3),
\end{aligned}
\end{equation}
\begin{equation}
\label{eq:2ptT}
\begin{aligned}
	\langle T_{zz}(z) T_{zz}(0) \rangle_\lambda &= \frac{c}{2z^4} + \frac{5\pi^2 \lambda^2 c^2}{6} \frac{1}{z^6 \oz^2}+ O(\lambda^3), \\
    \langle T_{zz}(z) T_{z\oz}(0) \rangle_\lambda &= - \frac{\pi^2\lambda^2 c^2}{3}\frac{1}{z^5\oz^3}+ O(\lambda^3), \\
    \langle T_{zz}(z) T_{\oz\oz}(0) \rangle_\lambda &= \frac{\pi^2 \lambda^2 c^2}{4} \frac{1}{z^4 \oz^4}+ O(\lambda^3), \\
    \langle T_{z\oz}(z) T_{z\oz}(0) \rangle_\lambda &= \frac{\pi^2 \lambda^2 c^2}{4}\frac{1}{z^4\oz^4}+ O(\lambda^3).
\end{aligned}
\end{equation}

 We can see there is no space-filling brane current $C$ at leading order in $\lambda$ from the vanishing of the two point function
 \begin{equation}
     \langle S_z(z) S_{\overline{z}}(0) \rangle_\lambda = 0,
 \end{equation}
 because as pointed out in \cite{Dumitrescu:2011iu}, $C$ appears in the two-point function as
\begin{equation}
    \langle S_z(z) S_{\overline{z}}(0) \rangle \sim \frac{C}{\overline{z}}.
\end{equation}
This is consistent with the existence of $\mR_{\overline{z}}$.

\subsection{Deformed three-point correlators}
Before we calculate the three-point correlators for the deformed $\mathcal{N} = (0, 2)$ SCFT, we first write down all the non-zero three-point correlators in the undeformed case. Reading off from the OPEs (\ref{OPEcomponents}), the undeformed three-point correlators are:
\begin{equation}\label{3pointSCFT}
\begin{aligned}
	\langle T_{zz}(x_1) T_{zz}(x_2) T_{zz}(x_3) \rangle_0 &= \frac{c}{z_{12}^2 z_{23}^2 z_{31}^2}, \\
    \langle T_{\oz\oz}(x_1) T_{\oz\oz}(x_2) T_{\oz\oz}(x_3) \rangle_0 &= \frac{c}{\oz_{12}^2\oz_{23}^2\oz_{31}^2}, \\
    \langle T_{zz}(x_1) S_z(z_2) \oS_z(z_3) \rangle_0&= \frac{c}{z_{12}^2 z_{31}^2 z_{31}}, \\
    \langle j_z(z_1) S_z(z_2) \oS_z(z_3) \rangle_0 &= \frac{2c}{3} \frac{1}{z_{12}z_{31}z_{23}^2}, \\
    \langle T_{zz}(z_1)j_z(z_2)j_z(z_3) \rangle_0&= \frac{c}{3} \frac{1}{z_{12}^2z_{31}^2}.
\end{aligned}
\end{equation}

Next, we consider the deformed three-point correlators $\langle O_1(z_1) O_2(z_2) O_3(z_3) \rangle_\lambda$ at $O(\lambda)$. The three-point correlators fall into two separate classes: 
\begin{enumerate}
    \item We could have one of $O_i$, say $O_1$, being an operator which vanishes in the SCFT, or at the superconformal point
    (e.g., $T_{z\oz}$, $j_{\oz}$, $\mR_{\oz}$),
    and the rest of them being the operators which survive in the
    SCFT (e.g., $T_{zz}$, $T_{\oz\oz}$, $\mR_{z}$).
    \item  We could have all three of them being the operators which survive in the
    SCFT.
\end{enumerate}

For the first case, at the first order in perturbation theory, we obtain a four-point correlator in the original SCFT because the first operator in the three-point correlator is a product of holomorphic and anti-holomorphic operators $T_{\oz\oz}$. The rest of the two operators must be the same holomorphic operator and $T_{\oz\oz}$ in order to have a non-vanishing result.

For the first case, correlators can be easily computed using superspace notation:
\begin{equation}
\begin{split}
    \langle \mR_{\oz}(Z_1) \mR_{z}(Z_2) T_{\oz\oz}(z_3) \rangle_\lambda & = - \pi \lambda \langle \mR_{z}(Z_1) \mR_{z}(Z_2) \rangle \langle T_{\oz\oz}(z_1) T_{\oz\oz}(z_3) \rangle  + O(\lambda^2) \\
    & = - \frac{\pi\lambda c^2}{6 Z_{12}^2 \oz_{31}^4}  + O(\lambda^2),
\end{split}
\end{equation}
and we collect the three-point functions in component fields at the end of this subsection.

For the second case, we obtain a five-point correlator in the undeformed SCFT because one of the operators has to be $T_{\oz\oz}$ which is anti-holomorphic. We must have another one or two operators being $T_{\oz\oz}$ to see a non-vanishing result since the five-point function factorizes into a product of correlation functions of holomorphic and anti-holomorphic operators. If the deformed three-point correlator contains two $T_{\oz\oz}$, then 
\begin{equation}
\begin{aligned}
    \langle O_1(x_1) T_{\oz\oz}(x_2) T_{\oz\oz}(x_3)\rangle_\lambda & = - \pi \lambda \int d^2x \langle O_1(x_1) T_{\oz\oz}(x_2) T_{\oz\oz}(x_3) T_{zz} (x) T_{\oz\oz}(x) \rangle_0  + O(\lambda^2)\\
    & = - \pi \lambda \int d^2x \langle O_1(x_1) T_{zz}(x) \rangle_0 \langle T_{\oz\oz}(x_2) T_{\oz\oz}(x_3) T_{\oz\oz}(x) \rangle_0  + O(\lambda^2).
\end{aligned}
\end{equation}
To calculate a non-zero result, we must set $O_1 = T_{zz}$ and obtain what \cite{Kraus:2018xrn} found :
\begin{equation}
\label{eq:3ptT}
    \langle T_{zz}(x_1) T_{\oz\oz}(x_2) T_{\oz\oz}(x_3) \rangle_\lambda = - \frac{\pi\lambda c^2}{3} \frac{1}{z_{12}^3}\frac{1}{\oz_{23}^5} - \frac{\pi\lambda c^2}{3}\frac{1}{z_{31}^3\oz_{23}^5}  + O(\lambda^2).
\end{equation}

If it contains only one of $T_{\oz\oz}$, then we have
\begin{equation}
\begin{aligned}
    \langle O_1(x_1) O_2(x_2) \oT_{\oz \oz}(x_3)\rangle_\lambda & = - \pi \lambda \int d^2x \langle O_1(x_1) O_2(x_2) \oT_{\oz\oz}(x_3) T_{zz}(x) \oT_{\oz \oz}(x) \rangle_0  + O(\lambda^2) \\
    & = - \pi \lambda \int d^2x \langle O_1(x_1) O_2(x_2) T_{zz}(x) \rangle_0 \langle \oT_{\oz \oz}(x_3) \oT_{\oz \oz}(x) \rangle_0  + O(\lambda^2).
\end{aligned}
\end{equation}
To obtain a non-zero result, we must have $\langle O_1(x_1) O_2(x_2) T(x) \rangle_0 \neq 0$. Using (\ref{3pointSCFT}) and the trace flow equation as systematically carried out in \cite{Kraus:2018xrn} for the deformed three-point correlators at $O(\lambda)$, we summarize all the non-vanishing three-point correlators:
\begin{equation}
\begin{aligned}
	\langle j_{\oz}(x_1) j_z(x_2) T_{\oz\oz}(x_3) \rangle_\lambda & = -\frac{\pi \lambda c^2}{6} \frac{1}{z_{12}^2 \oz_{31}^4} + O(\lambda^2), \\
    \langle S_{\oz}(x_1) \oS_z(x_2) T_{\oz\oz}(x_3) \rangle_\lambda &= -\frac{\pi \lambda c^2}{6} \frac{1}{z_{12}^3 \oz_{31}^4} + O(\lambda^2), \\
    \langle \oS_{\oz}(x_1) S_z(x_2) T_{\oz\oz}(x_3) \rangle_\lambda &= -\frac{\pi \lambda c^2}{6} \frac{1}{z_{12}^3 \oz_{31}^4} + O(\lambda^2), \\
    \langle j_z(x_1) j_z(x_2) T_{\oz\oz}(x_3) \rangle_\lambda &= -\frac{2\pi \lambda c^2}{9} \bigg(\frac{1}{\oz_{23}^3 z_{12}^3} + \frac{1}{z_{12}^3 \oz_{31}^3}\bigg) + O(\lambda^2), \\
    \langle S_z(x_1) \oS_z(x_2) T_{\oz\oz}(x_3) \rangle_\lambda &= - \frac{\pi \lambda c^2}{3}\bigg(\frac{1}{z_{12}^5\oz_{31}^3} + \frac{1}{\oz_{23}^3z_{12}^5}\bigg) + O(\lambda^2), \\
    \langle T_{zz}(x_1) T_{\oz\oz}(x_2) T_{\oz\oz}(x_3) \rangle_\lambda &= - \frac{\pi\lambda c^2}{3} \bigg(\frac{1}{z_{12}^3}\frac{1}{\oz_{23}^5} + \frac{1}{z_{31}^3\oz_{23}^5}\bigg) + O(\lambda^2).
\end{aligned}
\end{equation}

\subsection{Deformed \texorpdfstring{$n$}{}-point correlators of other operators}
We briefly comment on how to compute $n$-point correlators between $n$ generic supermultiplets using conformal perturbation theory. An analysis for the deformed $\mathcal{N} = (1,1)$ and $\mathcal{N} = (2,2)$ supersymmetric correlators were done by \cite{He:2019ahx}. Recall that $\mathcal{N} = (2,2)$ superspace are described by holomorphic and anti-holomorphic coordinates $(Z, \widetilde{Z}) = (z, \theta, \bar{\theta}, \bar{z}, \widetilde{\theta}, \bar{\widetilde{\theta}})$. To obtain  $\mathcal{N} = (0,2)$, we simply set these two fermionic coordinates to zero $\widetilde{\theta} = \bar{\widetilde{\theta}} = 0$ in $\mathcal{N} = (2,2)$ superspace. A generic $\mathcal{N} = (0,2)$ superfield is given by \cite{Kiritsis:1987np}:
\begin{equation}
    	\Phi_i(Z_i) = \phi_i(z_i) + \theta_i \overline{\psi}_i(z_i) + \otheta_i \psi_i(z_i) + \theta\otheta g_i(z_i),
\end{equation}
where $\phi_i(z), g_i(z_i)$ are complex scalar fields and $\psi_i (z_i)$ is a complex spinor. Therefore, at leading order in $\lambda$, the deformed $n$-point correlators are simply given by 
\begin{equation}
     \left \langle \prod^n_{i = 1} \Phi_i (Z_i, \overline{Z}_i) \right \rangle_\lambda =   -\lambda \int d^2z \int d^2\theta \left \langle \mS^{(0)}(Z)T_{\oz\oz}(\oz) \prod^n_{i=1}\Phi_i(Z_i,\oZ_i) \right \rangle_0,
\end{equation}
where 
\begin{equation}
    \int d^2\theta ~ \mathcal{S}^{(0)}(Z) = T_{zz} (z).
\end{equation}
This result can be conveniently acquired using the $\mathcal{N} = (2,2)$ results in \cite{He:2019ahx} by setting $\tilde{\theta} = \bar{\tilde{\theta}} = 0$.

\subsection{Renormalized correlators and universality}
An interesting question to ask is if one can define {\it renormalized} operators as in \cite{Cardy:2019qao,Kraus:2018xrn} while respecting supersymmetry. This proof is quite obvious in superfield formalism inspired from \cite{He:2019ahx}. The singular piece in the dimension regularization parameter $\epsilon$ for the deformed two-point correlator is
    \begin{equation}
        \langle \Phi_1(Z_1,\oz_1)\Phi_2(Z_2,\oz_2)\rangle_\lambda = - \frac{16\pi\lambda}{\epsilon}\bigg(\frac{2h}{Z_{12}} + Q_2 \frac{\theta_{12}\otheta_{12}}{z_{12}^2}\bigg)\frac{2\oh}{\oz_{12}}\langle \Phi_1(Z_1,\oz_1)\Phi_2(Z_2,\oz_2)\rangle_0,
    \end{equation}
    where the undeformed two-point correlator is 
    \begin{equation}
        \langle \Phi_1(Z_1,\oz_1)\Phi_2(Z_2,\oz_2)\rangle_0 = \frac{1}{Z_{12}^{2h}\oz_{12}^{2\oh}}e^{Q_2\frac{\theta_{12}\otheta_{12}}{Z_{12}}}\delta_{Q_1+Q_2,0}.
    \end{equation}
    Here $(h_1,\oh_1) = (h_2,\oh_2) = (h,\oh)$ are the scaling dimensions and $Q_1 + Q_2 = \oQ_1 + \oQ_2 = 0$ are the charges of $\Phi_1(Z_1, z_1)$ and $\Phi_2(Z_2, z_2)$.
    
    We define the {\it renormalized} superfield as
    \begin{equation}
        \Phi_R = \Phi - \frac{A \lambda}{\epsilon}\{D,\oD\}\partial_{\oz} \Phi = \Phi - \frac{2A\lambda}{\epsilon} \partial_z \partial_{\oz} \Phi
    \end{equation}
    in order to remove the above singular piece. In \cite{Kraus:2018xrn}, for a general operator $\mathcal{O}$ with scaling dimension $(h_{\mathcal{O}},\oh_{\mathcal{O}})$, we have 
    \begin{equation}
        \langle \mathcal{O}(x_1) \mathcal{O}^\dagger(x_2) T(x) \oT(x) \rangle_0 = h_{\mathcal{O}}\oh_{\mathcal{O}} \bigg(\frac{(x_1-x_2)^2}{(x-x_1)^2(x-x_2)^2}\bigg)^2 \langle \mathcal{O}(x_1) \mathcal{O}^\dagger(x_2)\rangle_0.
    \end{equation}
    So, the deformed two-point correlators differ only by an overall coefficient $h_{\mathcal{O}} \oh_{\mathcal{O}}$ and this matches the coefficients one obtains from acting on the undeformed two-point correlators with $\partial_z \partial_{\oz}$:
    \begin{equation}
        \partial_z\partial_{\oz} \frac{1}{z^{2h_{\mathcal{O}}}\oz^{2\oh_{\mathcal{O}}}} = \frac{4h_{\mathcal{O}}\oh_{\mathcal{O}}}{z^{2h_{\mathcal{O}}+1}\oz^{2\oh_{\mathcal{O}}+1}},
    \end{equation}
    which implies the coefficient $A$ is universal for every operator. This allows us to embed the $T\oT$ deformation into superfield formalism, and thus preserve supersymmetry. As a preview to the upcoming subsection, we will see that such a renormalization is absent for the $\mathcal{N} = (0, 2)$ chiral ring elements $\Phi$ which satisfy $\partial_{\oz} \Phi = 0$.
    
\subsection{Other deformed operators and chiral rings}
We will also comment on the $T\oT$ deformation for other operators. The essential results have been worked out by Cardy \cite{Cardy:2019qao} and the advantage here is since the deformation is independent of the operator's scaling dimension, one can easily write the deformation in superfield formalism for a given supermultiplet. In the upcoming subsections, we will provide a few comments on more operators, such as conserved currents and chiral ring elements in $\mathcal{N} = (0,2)$ theories.

\subsubsection{Deformed holomorphic currents}
Here we study holomorphic currents under the $T\oT$ deformation. In a $\mathcal{N} = (0,2)$ SCFT with a normalizable vacuum, the $\mathcal{N} = (0,2)$ holomorphic multiplet current $\mJ_z^A$ is constructed out of two copies of $\mathcal{N} = (0,1)$ currents $(\psi^A,j_z^A)$. For simplicity, we consider these currents to be abelian and work in a complex basis: $\mJ = \frac{1}{\sqrt{2}}\left(\mJ_1 + i \mJ_2\right)$. Thus, we have
\begin{equation}
\begin{aligned}
    \mJ_z^A &= \psi^A + i\sqrt{2}\otheta j^A_z - \theta\otheta \partial_z \psi^A, \\
    \omJ_z^A&= \opsi^A - i\sqrt{2}\theta \oj^A_z + \theta\otheta \partial_z \opsi^A,
\end{aligned}
\end{equation}
which are conserved
\begin{equation}
    \partial_z \mJ^A = \partial_z \omJ^A = 0, \quad  \oD \mJ = D \omJ = 0.
\end{equation}
The undeformed current OPEs are
\begin{equation}
\begin{aligned}
    \mJ_z^A(Z_1) \omJ_z^B(Z_2) &= \frac{k^{AB}}{Z_{12}}e^{-\frac{\theta_{12}\otheta_{12}}{Z_{12}}}= k^{AB}\bigg(\frac{1}{Z_{12}} - \frac{\theta_{12}\otheta_{12}}{Z_{12}^2}\bigg),
\end{aligned}
\end{equation}
where $k^{AB}$ are `t Hooft anomaly coefficients
\begin{equation}
    k^{AB}=\left\{\begin{array}{ll}
k_{R}^{AB} & \text { if } A, B \text { are both right-moving } \\
-k_{L}^{AB} & \text { if } A, B \text { are both left-moving }, \\
0 & \text { otherwise }
\end{array} \quad k=c_{R}-c_{L}\right.
\end{equation}

Under the $T\oT$ deformation, we need to accompany $\mJ_z^{(0)}$ with 
\begin{equation}
    \mJ_{\oz} = - \pi\lambda \mJ_{z}^{(0)} T_{\oz\oz} + O(\lambda^2)
\end{equation}
such that\footnote{As one can easily check $\oD \mJ_{\oz} = D \omJ_{\oz} = 0$ remains chiral.}
\begin{equation}
    \partial_{z} \mJ_{\oz} + \partial_{\oz} \mJ_{z} = 0.
\end{equation}
Then, just as before in the previous subsections, one can write down the deformed current two-point correlator
\begin{equation}
\begin{aligned}
    \left \langle  \mJ_{\overline{z}}^A(Z_1) \omJ_{\overline{z}}^B(Z_2) \right  \rangle_\lambda &= \pi^2 \lambda^2 \left \langle \mJ_z^A(Z_1) \omJ_z^B(Z_2)\right  \rangle_0 \left \langle \oT_{\oz \oz} (z_1) \oT_{\oz \oz} (z_2) \right \rangle_0+ O(\lambda^3) \\&= \pi^2 \lambda^2 \frac{k^{AB}}{\oz^4_{12}}\bigg(\frac{1}{Z_{12}} - \frac{\theta_{12}\otheta_{12}}{Z_{12}^2}\bigg) + O(\lambda^3).
\end{aligned}
\end{equation}

Likewise, how conservation equations were used to find the other $\mathcal{R}$ correlators in (\ref{otherRviaconservation}) in principle, one can perform the same trick for the other current correlators $ \left \langle  \mJ_z^A(Z_1) \mJ_z^B(Z_2) \right  \rangle_\lambda$ and $ \left \langle  \mJ_z^A(Z_1) \mJ_{\oz}^B(Z_2) \right  \rangle_\lambda$.

\subsubsection{Deformed two-dimensional \texorpdfstring{$\mathcal{N} = (0,2)$}{} chiral ring}
Another interesting question to analyze is the $T\oT$ deformed chiral ring elements for two-dimensional $\mathcal{N} = (0,2)$ SCFTs. These operators are superconformal primaries that are annihilated by half of the superconformal charges, $G^+_n$ or $G^-_n$, and saturate the unitarity bound $h = |q|/2$. Under the deformation, superconformal symmetry is broken and the full short representation splits into representations of the surviving supersymmetry. Hence, a natural question one can address is whether the resulting representations are chiral? To show whether half of the supercharges annihilate these deformed operators, one can compute OPE between the deformed supercurrent and the deformed chiral operators at leading order in $\lambda$. Let $\Phi^{(0)}$ be an element of the two-dimensional $\mathcal{N} = (0,2)$ chiral ring and is annihilated by $G^-_z$: $G_{-1/2}^-|\Phi^{(0)}\rangle = 0$. Then, we have the OPEs
\begin{equation}
\begin{aligned}
    G^-_z{}^{(0)}(z) \Phi(0) &\sim 0, \\ T_{\oz\oz}^{(0)}(z) \Phi(0) &\sim 0.
\end{aligned}
\end{equation}
Now by using the deformation equation, $G^-_{\oz} = - \pi \lambda G^-_{z}{}^{(0)} T_{\oz\oz}{}^{(0)} + O(\lambda^2)$, we find
\begin{equation}
    G^-_{\oz}(z) \Phi(0) = - \pi \lambda G^-_{z}{}^{(0)}(z) T_{\oz\oz}(z) \Phi(0)^{(0)} + O(\lambda^2) \sim O(\lambda)^2.
\end{equation}
From also using 
\begin{equation}
    \delta \Phi(0) = \pi \lambda \int_0^X d\oz' T_{\oz\oz}{}^{(0)}(z') \partial_z\Phi(0) + O(\lambda^2)
\end{equation}
where $X$ is an arbitrary reference point and
\begin{equation}
    \delta T_{zz}(z) = \pi \lambda \int_z^X d\oz' T_{\oz\oz}^{(0)}(z')\partial_z G^-_{z}{}^{(0)}(z) + O(\lambda^2),
\end{equation}
we obtain 
\begin{equation}
    \delta(G_{z}^-(z) \Phi(0)) = [\delta G_{z}^-(z)] \Phi^{(0)}(0) + G_{z}^-{}^{(0)}(z) \delta \Phi(0) \sim 0 + O(\lambda^2).
\end{equation}
As a result,
\begin{equation}
    G_{z}^-(z) \Phi(0) \sim G_{\overline{z}}^-(z) \Phi(0) \sim O(\lambda^2)
\end{equation}
which implies $\Phi(x)$ and its super-partner will form a chiral superfield. 

Another interesting perspective is to check the chiral ring relation under the $T\oT$ deformation. The deformed OPE coefficients for a general operator is derived in \cite{Cardy:2019qao}:
\begin{equation}
    \delta C^l{}_{mn}(x_1-x_2) = 2\pi \lambda \epsilon^{ab}\int_{x_1}^{x_2} T_{ai}(x'+\epsilon) \epsilon^{ij}dx_j' \partial_b C^l{}_{mn}(x_1-x_2),
\end{equation}
but since the OPE coefficients inside the chiral ring are constant, i.e., $\partial_{x^i} C^l{}_{mn}(x) = 0$, we must have $\delta C^l{}_{mn} = 0$. Therefore, the $\mathcal{N} = (0, 2)$ chiral ring relation is preserved under the deformation. 

\section{Deformation in two-dimensional \texorpdfstring{$\mathcal{N} = (2,2)$}{} SCFT}
\label{sec:(2,2)}
In this section, we study the deformed $\mathcal{S}$-multiplet for $\mathcal{N} = (2,2)$ SCFTs. The calculations for the $\mS$-multiplet are similar to those in $\mathcal{N} = (0,2)$ theories in the previous section. The main difference is that we must accompany the $T\overline{T}$ deformation with an additional improvement transformation to preserve one of the $U(1)_R$ symmetries in $\mathcal{N} = (2,2)$ theories. 

\subsection{A brief review of two-dimensional \texorpdfstring{$\mathcal{N} = (2, 2)$}{} \texorpdfstring{$\mathcal{S}$}{}-multiplets}
In this subsection, we briefly review two-dimensional $\mathcal{N} = (2,2)$ $\mathcal{S}$-multiplets following \cite{Dumitrescu:2011iu}. A generic $\mathcal{N} = (2,2)$ $S$-multiplet without conformal symmetry consists of two real superfields $\mS_{\pm\pm}$ together with chiral superfields $\chi_\pm$ and twisted chiral superfields $\mathcal{Y}_\pm$ satisfying the following constraints
\begin{equation}
    \overline{D}_{\pm} \mS_{\mp\mp} = \pm (\chi_\mp + \mathcal{Y}_\mp),
\end{equation}
where
\begin{equation}
\begin{array}{ll}
    \overline{D}_\pm \chi_\pm = 0, & D_\pm \mathcal{Y}_\pm = 0, \\
    \overline{D}_\pm \chi_\mp = \pm C^{(\pm)}, & \overline{D}_\pm \mathcal{Y}_\mp = \mp C^{(\pm)}, \\
    D_+\chi_- - \overline{D}_- \overline{\chi}_+ = k, & D_+ \mathcal{Y}_- + D_- \mathcal{Y}_+ = k'.
\end{array}
\end{equation}
Here $k,k'$ and $C^{(\pm)}$ are real and complex constants respectively. 

In components, the $\mathcal{S}$-multiplets are given by
\begin{equation}
\begin{aligned}
    \mathcal{S}_{\pm\pm} = & j_{\pm\pm} - i \theta^\pm S_{\pm\pm\pm} - i \theta^\mp ( S_{\mp\pm\pm} \mp 2 \sqrt{2} i \opsi_\pm) - i \otheta^\pm  S_{\pm\pm\pm} - i \otheta^\mp \left(\oS_{\mp\pm\pm} \pm 2 \sqrt{2} i \psi_\pm\right) \\
    &  - \theta^\pm \otheta^\pm T_{\pm\pm\pm\pm} + \theta^\mp \otheta^\mp \left(A \mp{ k + \frac{k'}{2} }\right) + i \theta^+ \theta^-  \oY_{\pm\pm} + i \otheta^+ \otheta^- Y_{\pm\pm} \\
    & \pm i \theta^+ \otheta^- \oG_{\pm\pm} \mp i \theta^- \otheta^+ G_{\pm\pm} \mp \frac{1}{2} \theta^+ \theta^- \otheta^\pm \partial_{\pm\pm} S_{\mp\pm\pm} \mp \frac{1}{2} \theta^+ \theta^- \otheta^\mp \partial_{\pm\pm}\left(S_{\pm\mp\mp} \pm 2 \sqrt{2} i \opsi_\mp\right) \\
    & \mp \frac{1}{2} \otheta^+ \otheta^- \theta^\pm \partial_{\pm\pm} \oS_{\mp\pm\pm} \mp \frac{1}{2} \otheta^+ \otheta^-\theta^\mp \partial_{\pm\pm} \left( \oS_{\pm\mp\mp} \mp 2 \sqrt{2}  i \psi_\mp\right) + {\frac{1}{4}} \theta^+ \theta^- \otheta^+ \otheta^- \partial_{\pm\pm}^2 j_{\mp\mp}
\end{aligned}
\end{equation}
and
\begin{equation}
\begin{aligned}
    \chi_+ &= - i \lambda_+(y) - i \theta^+ \oG_{++}(y) +  \theta^- \left(E(y) + {\frac{k}{2}}\right) + \otheta^- C^{(-)}  + \theta^+ \theta^- \partial_{++} \olambda_-(y)~,\\
    \chi_- &= - i \lambda_-(y) - \theta^+ \left(\overline{E}(y) - \frac{k}{2}\right) + i \theta^- G_{--}(y) - \otheta^+ C^{(+)}- \theta^+ \theta^- \partial_{--} \olambda_+(y),\\
    \lambda_\pm &= \pm\oS_{\mp\pm\pm} + \sqrt{2} i \psi_\pm,\\
    E &= \frac{1}{2} \left( T_{++--} - A \right)+ \frac{i}{4} \left( \partial_{++} j_{--} - \partial_{--} j_{++}\right),\\
    \partial_{++} G_{--} &= \partial_{--} G_{++}, \\
    y^{\pm\pm} &= x^{\pm\pm} + 4 i  \theta^\pm\otheta^\pm,
\end{aligned}
\end{equation}

\begin{equation}
\begin{aligned}
    \mathcal{Y}_+ &= \sqrt{2} \psi_+ (\overline{\tilde{y}}) +  \theta^- \left( F(\overline{\tilde{y}}) + {\frac{k'}{2}}\right) - i \otheta^+ Y_{++}(\overline{\tilde{y}}) - \otheta^- C^{(-)}+ \sqrt{2} i \theta^- \otheta^+ \partial_{++} \psi_-(\overline {\tilde{y}}), \\
    \mathcal{Y}_- &= \sqrt{2} \psi_-(\tilde{y}) -  \theta^+ \left( F(\tilde{y}) - \frac{k'}{2}\right) + \otheta^+ C^{(+)}- i \otheta^- Y_{--}(\tilde{y}) + \sqrt{2} i \theta^+ \otheta^- \partial_{--} \psi_+ (\tilde{y}), \\
    F &= - \frac{1}{2} \left( T_{++--} + A\right) - \frac{i}{4} \left(\partial_{++} j_{--} + \partial_{--} j_{++}\right),\\
    \partial_{++} Y_{--} &= \partial_{--} Y_{++},\\
    \tilde{y}^{\pm\pm} &= x^{\pm\pm} \pm 4 i \theta^\pm \otheta^\pm.
\end{aligned}
\end{equation}
It is important to identify the conserved brane currents in the generic $\mS$-multiplet: $\mp i \oY_{\pm\pm}$ and $\pm i\oG_{\pm\pm}$ are the zero-brane currents that give rise to the central charges $Z$ and $\tilde{Z}$, while the constants $C^{\pm}$ and $k-k'$ are the space-filling brane currents which can lead to partial supersymmetry-breaking. 

The $\mS$-multiplet can be modified by an improvement transformation by a real superfield $U$:
\begin{equation}
\begin{aligned}
    \mS_{\pm\pm} &\rightarrow \mS_{\pm\pm} + [D_\pm,\overline{D}_\pm] U, \\
    \chi_{\pm} &\rightarrow \chi_\pm - \overline{D}_+ \overline{D}_- D_\pm U, \\
    \mathcal{Y}_\pm &\rightarrow \mathcal{Y}_\pm - D_\pm \overline{D}_+ \overline{D}_- U.
\end{aligned}
\end{equation}

Like in the $\mathcal{N} = (0,2)$ case, we are interested in the various possibilities where the $\mS$-multiplet can be improved to a smaller multiplet: 
\begin{enumerate}
    \item If $k = C^{(\pm)} = 0$ and there is a well-defined $U$ such that $\chi_\pm = \overline{D}_+\overline{D}_- D_\pm U$, then an improvement transformation can be used to set $\chi_\pm = 0$. The resulting multiplet is the FZ-multiplet, whose bottom component is a conserved axial current.
    \item If instead $k' = C^{(\pm)} = 0$ and there is a well-defined $U$ such that $\mathcal{Y}_\pm = D_\pm \overline{D}_+ \overline{D}_- U$, then a improvement transformation will set $\mathcal{Y}= 0$. This leads to the $\mR$-multiplet, whose bottom component is a conserved vector current. Notice that the FZ- and $\mR$-multiplets are related by the following mirror automorphisms:
\begin{equation}
    \mS_{\pm\pm}\leftrightarrow \pm \mS_{\pm\pm}, \,\,\, \chi_+ \leftrightarrow \overline{\mathcal{Y}}_+, \,\,\, \chi_- \leftrightarrow -\mathcal{Y}_-,\,\,\,
    k \leftrightarrow - k', \,\,\, C^{(\pm)} \leftrightarrow \overline{C}^{(\pm)}.
\end{equation}
\item If $k = k' = C^{(\pm)} = 0$ and both $\chi_\pm$ and $\mathcal{Y}_{\pm}$ can be removed by an improvement transformation, then the theory is superconformal.
\end{enumerate}

\subsection{Deformation of the generic \texorpdfstring{$\mathcal{S}$}{}-multiplet in \texorpdfstring{$\mathcal{N} = (2,2)$}{} SCFT}
We derive the deformed $\mS$-multiplet by supersymmetrizing the flow equation (\ref{eq:traceflow}). The trick is that we can apply the supercharge $Q$ on both sides of the deformation equation. We have to be careful since the supercharge $Q$ in the deformed theory would be different from the supercharge $Q^{(0)}$ in the seed theory by $O(\lambda)$ corrections. Since $T_{z\oz}$ is already of the first order in $\lambda$, at leading order of the perturbation theory, the extra correction of $\lambda$ will not contribute. We have
\begin{equation}
    [Q,T_{z\oz}] = [Q^{(0)}, - \pi \lambda T_{zz} T_{\oz\oz}] + O(\lambda^2).
\end{equation}
One might worry that supersymmetry acts on operators which do not vanish at the superconformal point, such as $T_{zz}$, will receive contribution like $[\lambda Q^{(1)}, T_{zz}^{(0)}] $; however, all the leading order contributions can be acquired by acting the supercharge on the operators which do vanish at the superconformal point together with the Ward identity. Thus, we do not have to worry about such a contribution. Using this method, we can determine the deformed $\mathcal{S}$-multiplet. The collected results are rather extensive and we refer the reader to appendix \ref{App:22}. 

The advantage of this calculation is that neither $Y_{\pm\pm}$ or $G_{\pm\pm}$ vanishes so that we acquire a generic $\mS$-multiplet with no conserved $R$-currents. In particular, the deformation of the bottom components $j_{\pm\pm} \subset \mS_{\pm\pm}$ are given by 

\begin{equation}
\begin{aligned}
\label{J}
    & \partial_{++} j_{--} = - \frac{\pi\lambda}{16} \partial_{--} j_{--}^{(0)} T_{++++}^{(0)} + O(\lambda^2) , \\
    & \partial_{--} j_{++} = - \frac{\pi\lambda}{16}\partial_{++}j_{++}^{(0)} T_{----}^{(0)} + O(\lambda^2).
\end{aligned}
\end{equation}
The above result clearly indicates that neither the vector current nor axial current are conserved.

However, the appearance of the total derivative on the R.H.S. suggests it is possible to consider an improvement transformation to acquire a conserved $U(1)_A$ current or $U(1)_V$ current. Supersymmetrizing such improvement transformation leads to either the FZ-multiplet or $\mathcal{R}$-multiplet. For instance, one can consider the improvement transformation which acts on $j_{\pm\pm}$ via
\begin{equation}
    j_{++} \rightarrow j_{++} - \frac{\pi\lambda}{16} j_{--}^{(0)} T_{++++}^{(0)}, \,\,\, j_{--} \rightarrow  j_{--} - \frac{\pi\lambda}{16} j_{++}^{(0)} T_{----}^{(0)}.
\end{equation}
As one can check, after this improvement transformation, the vector current is now conserved at the leading order of the perturbation theory in $\lambda$:
\begin{equation}
    \partial_{++} j_{--} + \partial_{--} j_{++} = O(\lambda^2).
\end{equation}

Using $\mathcal{N} = (2,2)$ supersymmetry, one can find the improvement transformation for the result of the components in the $\mS$-multiplets and verify that the twisted chiral superfields $\mathcal{Y}_{\pm}$ indeed vanish. A new phenomenon is that this improvement transformation will lead to an additional correction to the trace flow equations: 
\begin{equation}
    T_{++--} = \frac{\pi\lambda}{16}T_{----}^{(0)}T_{++++}^{(0)} + \frac{\pi\lambda}{64}\partial_{--}j_{--}^{(0)}\partial_{++}j_{++}^{(0)} + O(\lambda^2).
\end{equation}

Notice that central currents $Y_{\pm\pm}$ and $G_{\pm\pm}$ will be generated under the $T\oT$ deformation. Even after the improvement transformation, only one of them can be removed and the one left takes form of a total derivative. For instance, if we choose to remove $Y_{\pm\pm}$ by an improvement transformation, we will find:
\begin{equation}
\begin{aligned}
    G_{++} &= -\frac{\pi\lambda}{16} \partial_{++} \oS_{+++}^{(0)}S_{---}^{(0)} + O(\lambda^2), \\
    G_{--} &= -\frac{\pi\lambda}{16} \oS_{+++}^{(0)} \partial_{--} S_{---}^{(0)} + O(\lambda^2), \\
    \oG_{++} &= \frac{\pi\lambda}{16} \partial_{++} S_{+++}^{(0)} \oS_{---}^{(0)} + O(\lambda^2), \\
    \oG_{--} &= \frac{\pi\lambda}{16} S_{+++}^{(0)} \partial_{--} \oS_{---}^{(0)} + O(\lambda^2).
\end{aligned}
\end{equation}
Being total derivatives itself does not suggest $G_{\pm\pm}$ leads to trivial charge, but rather it suggests one must understand non-perturbative effects of the $T\oT$ deformation to fully understand the perturbative $\mS$-multiplet for the deformed theory. Similar scenarios occur, such as calculating the instanton number in four-dimensional gauge theory from integrating the total derivative $\frac{1}{8\pi^2}\operatorname{Tr} F\wedge F$.

It might be tempting to conclude that either chiral ring or twisted chiral ring will cease to exist in the deformed theory. However, there is a caveat. There is no guarantee that all the would-be chiral ring or twisted chiral ring elements would actually be charged under the non-zero central current preventing one to simply draw such a conclusion. We believe understanding the non-perturbative effect of the $T\oT$ deformation and perhaps a model dependent analysis are required to determine the ultimate fate of the chiral ring and twisted chiral ring in the deformed theory.

\section{Deformed S-matrices in \texorpdfstring{$\mathcal{N} = (2, 2)$}{} integrable theories and indices}
\label{sec:S-matrix}
In the previous section, we studied the $T\oT$ deformation for $\mathcal{N} = (2,2)$ SCFTs and our results are perturbative in $\lambda$. The $T\oT$ deformation can be defined exactly for integrable models \cite{Cavaglia:2016oda, Smirnov:2016lqw, Rosenhaus:2019utc, Caetano:2020ofu, Cardy:2020olv, Jiang:2020nnb} and it is tempting to study integrable QFTs in order to make some exact statements via TBA.

Cavagli\`{a} et al. \cite{Cavaglia:2016oda} studied the $T\oT$ deformation for the well-known \textit{non-supersymmetric} sine-Gordon model using non-linear integral equations (NLIE). This is a single NLIE (for one particle/soliton), and by changing the integration contours, one can access different excited states. From this, one can derive a flow equation of the deformed energy $E$ for any state. 

However, for \textit{supersymmetric} integrable theories, most of the theories do not have such a simple description as NLIE. The most convenient method that is readily available is TBA and has been used in several instances to study integrable supersymmetric theories, such as in \cite{Fendley:1991ve,Fendley:1991xn,Fendley:1992dm,Fendley:1993pi,Cecotti:1992qh}. Unfortunately, rather than just a single integrable equation, the TBA system usually contains several coupled integrable equations.\footnote{There are hybrid approaches combining the TBA and NLIE. These approaches have been carried out in \cite{Hegedus:2005bg, Suzuki:2011dj, Balog:2012zt}, yet the NLIEs only replace half of the many coupled integral equations.} Another disadvantage of the TBA is that it is non-trivial to access generic states. Luckily, in \cite{Fendley:1991ve,Fendley:1991xn,Fendley:1992dm,Fendley:1993pi,Cecotti:1992qh}, supersymmetric observables such as the Witten index, CFIV index and elliptic genus have been calculated using TBA and we will adopt techniques such as NLIE and TBA to compute these observables under the $T\oT$ deformation. 

In this section, we will study $\mathcal{N} = (2, 2)$ Landau-Ginzburg models with superpotential
\begin{equation}
	W (X,\beta) = \frac{X^{n+1}}{n+1} - \beta X
\end{equation}
under the $T\oT$ deformation.

These LG models are already known to be integrable \cite{Fendley:1991ve, Fendley:1993pi} and we will study various physical aspects under the deformation following the strategies mentioned above after briefly reviewing the TBA system for the undeformed theory. Additionally, we will perform a TBA analysis for the $T\oT$ deformed $\mathcal{N} = (1, 1)$ integrable models and explicitly show that they are directly related to $\mathcal{N} = (2, 2)$ integrable models via Melzer's folding trick \cite{Melzer:1994qp}.  

\subsection{Review of TBA system for Landau-Ginzburg models}
In this subsection, we will provide a brief review on acquiring TBA for undeformed LG models. We skip detailed derivations and refer the reader to pioneering work by Fendley and Intriligator \cite{Fendley:1991ve,Fendley:1992dm}.

These LG models have $n$ supersymmetric vacua given by the $n$ solutions from $X^n = \beta$. There are BPS solitons which interpolate between different vacua. For a soliton connecting the vacuum $X_i$ and $X_f = e^{2\pi i r/n} X_i$, it has mass
\begin{equation}
    m_r = M \sin(r\mu), \,\,\, r = 1,\dots, n-1,
\end{equation}
where $M = \frac{2n}{n+1}$ and $\mu = \pi/n$. For each mass $m_r$, there are a pair of solitons $(u_r,d_r)$ related by $\mathcal{N} = (2,2)$ supersymmetry. 

LG models' S-matrices are diagonal under $r,s$ labels with the incoming and outgoing states being type-$r,s$ solitons. The S-matrix $S_{r,s} (\theta)$ is
\begin{equation}\label{LGSmatrix}
	\bordermatrix{&d_s u_r&u_s d_r\cr
u_r d_s&b_{r,s}(\theta)&{\tilde c}_{r,s}(\theta)\cr d_r u_s & c_{r,s}(\theta) &{\tilde b}_{r,s}(\theta)\cr} \qquad \qquad \bordermatrix{&u_s u_r &d_s d_r\cr
u_r u_s&a_{r,s}(\theta)&0\cr d_r d_s &0&{\tilde a}_{r,s}(\theta)\cr},
\end{equation}
where the S-matrix elements are
\begin{equation}
\label{eq:S}
\begin{array}{ll}
    a_{r,s}(\theta) = Z_{r,s}(\theta) \sinh\left(\frac{\theta}{2} + \frac{i\mu}{2}(r+s)\right), & \tilde{a}_{r,s}(\theta) = - Z_{r,s}(\theta) \sinh\left(\frac{\theta}{2}-\frac{i\mu}{2}(r+s)\right), \\
    b_{r,s}(\theta) = Z_{r,s}(\theta)\sinh\left(\frac{\theta}{2}+\frac{i\mu}{2}(s-r)\right), & \tilde{b}_{r,s}(\theta) = Z_{r,s}(\theta)\sinh\left(\frac{\theta}{2}+\frac{i\mu}{2}(r-s)\right), \\
    c_{r,s}(\theta) = Z_{r,s}(\theta) i e^{i\mu(r-s)/2}\left(\sin(r\mu)\sin(s\mu)\right)^{\frac{1}{2}}, & \tilde{c}_{r,s}(\theta) = Z_{r,s}(\theta)ie^{i\mu(s-r)/2}(\sin(r\mu)\sin(s\mu))^{\frac{1}{2}}
\end{array}
\end{equation}
and satisfy the condition
\begin{equation}
    a_{r,s}(\theta)\tilde{a}_{r,s}(\theta) + b_{r,s}(\theta)\tilde{b}_{r,s}(\theta) - c_{r,s}(\theta)\tilde{c}_{r,s}(\theta) = 0.
\end{equation}
The specific form of $Z_{r,s}(\theta)$ is unimportant in our analysis for the deformed theory and is thus omitted. 

The basic idea of TBA is to place the system on a torus of length $L$ (in the longitude direction) at temperature $T\equiv 1/R$. Then, working in the grand canonical ensemble, one finds the energy spectrum and the particular filling of the energy levels which minimize the free energy in the thermodynamic limit $L \rightarrow \infty$. 

Now, consider $N$ particles with rapidity $\theta_i, i = 1,\dots, N$ on a circle $S^1(R)$. The allowed wave function must be invariant under the transformation which brings the particle with rapidity $\theta_k$ around the circle and back to its original position. In other words, the wave function obeys the Yang equation (see e.g., \cite{zamolodchikov1991thermodynamic})
\begin{equation}
\label{singlevalued}
    e^{im_k \sinh\theta_k L} T(\theta_k|\theta_k,\cdots,\theta_N,\theta_1,\cdots,\theta_{k-1})\psi = \psi,
\end{equation}
where $T(\theta_k |\theta_{k+1}, \cdots, \theta_{k-1})$ is the transfer matrix. The transfer matrix's components can be expressed in terms of the S-matrix:
\begin{equation}
    \left(T_{a b}(\theta)\right)_{c_{i}}^{d_{i}} \equiv \sum_{k_1, \cdots, k_{N-1}} S_{a c_{1}}^{d_{1} k_{1}}\left(\theta-\theta_{1}\right) S_{k_{1} c_{2}}^{d_{2} k_{2}}\left(\theta-\theta_{2}\right) \cdots S_{k_{N-1} c_{N}}^{d_{N} b}\left(\theta-\theta_{N}\right).
\end{equation}

Once we obtain the transfer matrix's eigenvalues, the single-valuedness condition (\ref{singlevalued}) will lead to constraint equations. Minimizing the free energy under these constraints from (\ref{singlevalued}) provides the following TBA equations
\begin{equation}
    \epsilon_a(\theta) = m_a R\cosh(\theta) - \sum_b \int \frac{d\theta'}{2\pi} \phi_{ab}(\theta-\theta') \ln(1+e^{-\epsilon_b(\theta')}),
\end{equation}
where $\phi_{ab}(\theta)$ is a kernel and the distribution of type-$a$ solitons in the thermodynamic limit is $\rho_a(\theta) = \ln(1+e^{-\epsilon_a(\theta)})$. Therefore, the ground state energy of the system can be written as
\begin{equation}
    E(R) = - \sum_a \frac{m_a}{2\pi} \int d\theta \cosh(\theta) \ln(1+e^{-\epsilon_a(\theta)})
\end{equation}
and the ground state momentum $P(R)$ can be computed by replacing $\cosh\theta$ with $\sinh\theta$. Due to the symmetry in $\epsilon(\theta)=\epsilon(-\theta)$, one can show $P(R) \equiv 0$ and it is consistent with the expectation that the ground state has vanishing momentum. 

Solving for the transfer matrix's eigenvalues is usually a difficult task. For the special case where the S-matrix is diagonal, the only process are two type $a,b$ solitons scattering into two type $a,b$ solitons:
\begin{equation}
    S_{ab}^{cd}(\theta) = S_{ab}(\theta) \delta_{a}^c \delta_{b}^d.
\end{equation}
This implies that the transfer matrix is also diagonal and the TBA system then greatly simplifies. First, each $\epsilon_a (\theta)$ corresponds to a physical soliton of mass $m_a$. The kernel $\phi_{ab}$ also has a simple expression in terms of the S-matrix:
\begin{equation}
    \phi_{ab}(\theta) = - i \frac{\partial \ln S_{ab}(\theta)}{\partial \theta}.
\end{equation}

However, in (\ref{LGSmatrix}), the S-matrix for LG models are not diagonal. Fortuitously, the S-matrix is diagonal for the type-$r,s$ soliton. The S-matrix is of the 6-vertex model form and its eigenvalues can be solved by the so-called algebraic Bethe ansatz (ABA). From computing the eigenvalues of the S-matrix for fixed $r,s$, we construct the transfer matrix's eigenvalues. 

A detailed discussion on how ABA works can be found in \cite{Fendley:1991ve}'s appendix, but here we will only cite results relevant to our discussion. As we will see later, this general result for the 6-vertex model applies to the $T\oT$ deformed theory with no additional work required. 

Consider the undeformed S-matrix
\begin{equation}
	\bordermatrix{&u\overline{u}&d\overline{d}\cr
u\overline{u}&c&{b}\cr d\overline{d} & b &{c}\cr} \qquad \qquad \bordermatrix{&u \overline{d} & d \overline{u}\cr
d\overline{u}&0&a\cr u\overline{d} &a&{0}\cr},
\end{equation}
where its matrix elements obey
\begin{equation}
    a(\theta)\tilde{a}(\theta) + b(\theta)\tilde{b}(\theta) - c(\theta)\tilde{c}(\theta) = 0.
\end{equation}
The eigenvalues are
\begin{equation}
    \lambda(\theta; y) = \prod_{r=1}^m \frac{a(\theta -y_*)}{b(\theta -y_*)}\bigg[\prod_{i=1}^Nb(\theta -\theta_i)+(-1)^m\prod_{i=1}^N{\tilde a}(\theta -\theta_i)\bigg],
\end{equation}
where $y_r$ and $m$ are the solutions of 
\begin{equation}
    \prod_{i=1}^N \frac{b(y_*-\theta_i)}{\tilde{a}(y_*-\theta_i)} = (-1)^{m+1}.
\end{equation}

Since the S-matrix is diagonal with type-$r$ soliton, the transfer matrix's eigenvalues $\Lambda_r$ for bringing a type-$r$ soliton around the cycle and passing all other solitons would simply be a product of the 6-vertex model's eigenvalues corresponding to type-$r,s$ solitons. 

Due to the complexity from the non-diagonal S-matrix, the TBA system not only has $\epsilon_a(\theta)$ for each  type-$a$ soliton with mass $m_a$, but also contains two additional $\epsilon_{l}(\theta)$ with $l=0,\overline{0}$ and $m_l = 0$. The integral equations of these TBA systems are
\begin{equation}
\begin{aligned}
    \epsilon_l(\theta) &= - \sum_r \int \frac{d\theta'}{2\pi} \phi_{lr}(\theta-\theta') \text{ln}(1+e^{-\epsilon_r(\theta')}), \\
    \epsilon_s(\theta) &= m_s R\cosh(\theta) - \sum_{B} \int \frac{d\theta'}{2\pi} \phi_{sB}(\theta-\theta', 0) \text{ln}(1+e^{-\epsilon_B(\theta')}),
\end{aligned}
\end{equation}
where $B\in \{l,r\}$. 

We refer the readers to \cite{Fendley:1992dm} for detailed derivations of the kernels. The first kernel is
\begin{equation} \label{ker1}
  \phi_{lr}(\theta) = \frac{\sin(r\mu)}{\cosh(\theta) - a_l \cos(r\mu)}
\end{equation}
which does not depend on $Z_{r,s}(\theta)$ from the S-matrix \eqref{eq:S}. 

The second kernel is
\begin{equation} \label{ker2}
\begin{aligned}
   \phi_{rs}(\theta) &= \int \frac{dt}{2\pi} e^{it\theta} \bigg(\delta_{rs} - 2 \frac{\cosh\mu t \sinh(\pi-r\mu)t \sinh s\mu t}{\sinh \pi t \sinh \mu t}\bigg) \\
   &= \text{Im} \frac{d}{d\theta} \ln Z_{r,s}(\theta) + \dots,
\end{aligned}
\end{equation}
where $\dots$ denote the terms which are independent of $Z_{r,s}(\theta)$.

As we will see shortly in the following subsection, these two facts alone allow us to determine the TBA system for $T\oT$ deformed LG models. 

\subsection{TBA for the \texorpdfstring{$T\oT$}{} deformed LG models}

Following \cite{Cavaglia:2016oda}, the $T\oT$ deformation modifies the S-matrix by a Castillejo-Dalitz-Dyson (CDD) factor
\begin{equation}
    S^{kl}_{ij} (\theta, \lambda) = S^{kl}_{ij} (\theta,0) \Phi_{ij} (\theta, \lambda),
\end{equation}
where
\begin{equation}
    \Phi_{ij}(\theta,\lambda) = e^{i\lambda m_i m_j \sinh\theta}.
\end{equation}

As one can check, the deformed S-matrix satisfies all the standard constraints such as crossing symmetry, unitarity and Yang-Baxter equation.

Since the $T\oT$ deformation does not change the property that the S-matrix is diagonal in the type $r,s$ solitons, we can derive the TBA system following the same strategies used in the previous subsection. 

The first step is to derive the deformed eigenvalues of the 6-vertex model given by the S-matrix.

Recall from \cite{Fendley:1992dm}, the S-matrix $S_{r,s}(\theta)$ is
\begin{equation}
	\bordermatrix{&d_s u_r&u_s d_r\cr
u_r d_s&b_{r,s}(\theta)&{\tilde c}_{r,s}(\theta)\cr d_r u_s & c_{r,s}(\theta) &{\tilde b}_{r,s}(\theta)\cr} \qquad \qquad \bordermatrix{&u_s u_r &d_s d_r\cr
u_r u_s&a_{r,s}(\theta)&0\cr d_r d_s &0&{\tilde a}_{r,s}(\theta)\cr}
\end{equation}
and satisfies the constraint
\begin{equation}
    a_{r,s}(\theta)\tilde{a}_{r,s}(\theta) + b_{r,s}(\theta)\tilde{b}_{r,s}(\theta) - c_{r,s}(\theta)\tilde{c}_{r,s}(\theta) = 0.
\end{equation}

Due to $\mathcal{N} = (2,2)$ supersymmetry, $u_r$ and $d_s$ have the same mass and the $T\oT$ deformation modifies $S_{r,s}(\theta)$ by an overall phase factor $e^{i\lambda m_r m_s \sinh\theta}$. The incoming states always contain a type-$r$ soliton and a type-$s$ soliton. Thus, the constraint is still satisfied and, the general result will hold for the deformed theory!

As a result, all the derivations in the undeformed theory should easily go through for the deformed theory. The TBA system is obtained by absorbing the phase factor $e^{i\lambda m_r m_s \sinh\theta}$ into $Z_{r,s}(\theta)$, where we replace every $Z_{r,s}(\theta)$ with $e^{i\lambda m_r m_s \sinh\theta}Z_{r,s}(\theta)$. The deformed kernels are written as
\begin{equation}
\begin{aligned}
    \phi_{r,l}(\theta,\lambda) &= \phi_{r,l}(\theta,0), \\
    \phi_{r,s}(\theta,\lambda) &= \phi_{r,s}(\theta,0) + \lambda m_r m_s \cosh(\theta)
\end{aligned}
\end{equation}
and implies the integral equations take the following forms
\begin{equation}
\label{N=2TBA}
\begin{aligned}
    \epsilon_l(\theta) &= - \sum_r \int \frac{d\theta'}{2\pi} \phi_{lr}(\theta-\theta') \text{ln}(1+e^{-\epsilon_r(\theta')}), \\
    \epsilon_s(\theta, \lambda) &= m_s R \cosh(\theta) - \sum_r \lambda m_r m_s \int \frac{d\theta'}{2\pi} \big[\cosh(\theta)\cosh(\theta') + \sinh(\theta)\sinh(\theta') \big]  \text{ln}(1+e^{-\epsilon_r(\theta')}) \\
    &\quad - \sum_r \int \frac{d\theta'}{2\pi} \phi_{sr}(\theta - \theta') \text{ln}(1+e^{-\epsilon_r(\theta')}) - \sum_l \int \frac{d\theta'}{2\pi} \phi_{rl}(\theta-\theta')\text{ln}(1+e^{-\epsilon_l(\theta')}) \\
    &= m_s (R+\lambda E(R))\cosh(\theta) - \sum_{B} \int \frac{d\theta'}{2\pi} \phi_{sB}(\theta-\theta', 0) \text{ln}(1+e^{-\epsilon_B(\theta')}),
\end{aligned}
\end{equation}
where we have used
\begin{equation}
\begin{aligned}
    E(R) &= - \sum_r \frac{m_r}{2\pi} \int d\theta \cosh(\theta) \text{ln}(1+e^{-\epsilon_r(\theta)}), \\
    P(R) &= - \sum_r \frac{m_r}{2\pi} \int d\theta \sinh(\theta) \text{ln}(1+e^{-\epsilon_r(\theta)}) = 0.
\end{aligned}
\end{equation}

As previously mentioned, we arrive at the same result derived in \cite{Cavaglia:2016oda} for the non-supersymmetric sine-Gordon theory. The effect of the $T\oT$ deformation on the ground state energy shifts the radius $R$ by $R+\lambda E(R)$. Therefore, the diffeomorphism symmetry between the undeformed and deformed ground state energies is
\begin{equation}
    E(R, \lambda) = E(R+\lambda E(R,\lambda), 0),
\end{equation}
which famously obeys the inviscid Burgers' equation\footnote{Also called the {\it nonlinear advection equation}. Technically it is a \textit{quasilinear} equation, meaning that the PDE can be written in the form $a(x,y,u)u_x+b(x,y,u)u_y=c(x,y,u)$, not truly nonlinear.}
\begin{equation}
\label{eq:inviscid}
    \partial_\lambda E(R,\lambda) - E(R,\lambda) \partial_R E(R,\lambda) = 0.
\end{equation}
Generically, $E(R,0)$ can be solve numerically from the integral equation. However, in the UV limit $R\rightarrow 0$, we expect $E(R,0)$ to behave the same as the CFT \cite{Fendley:1991ve,Fendley:1992dm}
\begin{equation}
    E(R,0) \simeq - \frac{\pi c}{6R}, \quad R \rightarrow 0.
\end{equation}
It is tempting to solve $E(R,\lambda)$ near the $R = 0$ limit\footnote{For general solutions with nonzero momenta, see for example section 3 of \cite{Beratto:2019bap}.}, assuming the initial condition at $\lambda=0$ above:
\begin{equation}
\label{fullE(R,lambda)}
    E(R,\lambda) \simeq \frac{R}{2\lambda}\bigg(\sqrt{1-\frac{2\pi \lambda c}{3R^2}} - 1\bigg), \quad R \rightarrow 0.
\end{equation}
However, this leads to a double expansion in $R$ and $\lambda$:
\begin{equation}
    E(R,\lambda) = \bigg(-\frac{\pi c}{6R^1} + O\left(\frac{1}{R^0}\right)\bigg) + \lambda \left(-\frac{\pi^2 c^2}{36R^3} + O\left(\frac{1}{R^2}\right)\right) + \lambda^2 \left(-\frac{c^3\pi^3}{108R^5} + O\left(\frac{1}{R^4}\right)\right) + \cdots,
\end{equation}
where the higher order corrections in $O(\cdots)$ are coming from the subleading terms in the $E(R,0)$ expansion. If $\lambda$ is finite, we notice that higher order terms in $\lambda$ will be more dominant in the $R\rightarrow 0$ limit. As a result, \eqref{fullE(R,lambda)} is only true in the following double scaling limit between $\lambda$ and $R$:
\begin{equation}
\label{fullE(R,lambda)d}
    E(R,\lambda) \simeq \frac{R}{2\lambda}\bigg(\sqrt{1-\frac{2\pi \lambda c}{3R^2}} - 1\bigg), \quad R \rightarrow 0, \quad \lambda \rightarrow 0, \quad \frac{\lambda}{R^2} \sim O(1).
\end{equation}

We define the type-$r$ soliton's energy as
\begin{equation}
    E_r(R) = \frac{m_r}{2\pi} \int d\theta \cosh(\theta) \text{ln}(1+e^{-\epsilon_r(\theta)}).
\end{equation}

The physical interpretation of the individual $E_r(R)$ is clear -- we consider a grand canonical ensemble of solitons and derive a distribution for them which minimizes the Gibbs free energy. $E_r(R)$ is the energy contribution from a particular soliton type $r$. Yet, given this interpretation of $E(R)$ as the central charge $c$, one can ask the if interpreting $E_r(R)$ as a central charge $c_r$ makes sense for the type-$r$ soliton sub-theory such that $c = \sum_r c_r$. Although this interpretation sounds rather tempting, we comment it is false because solitons of different types interact with each other.

Therefore, $E_r(R,\lambda)$ are solutions to the one-dimensional coupled inviscid Burgers’ equations
\begin{equation}
\label{solveBurger}
    \partial_\lambda E_r (R,\lambda) - \bigg(\sum^n_{j=1} E_j(R,\lambda)\bigg) \partial_R E_r(R,\lambda) = 0.
\end{equation}

For two solitons, we have the following system of differential equations
\begin{equation}
\label{eq:Burgers}
    \begin{cases}
        \partial_\lambda E_1(R,\lambda) = \left( E_1(R,\lambda) + E_2 (R,\lambda)\right) \partial_R E_1 (R,\lambda), \\
       \partial_\lambda E_2(R,\lambda) = \left( E_1(R,\lambda) + E_2 (R,\lambda)\right) \partial_R E_2 (R,\lambda).
    \end{cases}
\end{equation}

Note that this is different from the full-fledged (with diffusion terms) {\it one}-dimensional coupled Burgers' equations for $u(x,t)$ and $v(x,t)$ (derived in a geophysical context of bidispeprsive sedimentation \cite{esipov1995coupled}), which has the following form:
\begin{equation}
\label{moregeneralburger}
    \begin{cases}
        \displaystyle{\frac{\partial u}{\partial t}+\delta\frac{\partial^2 u}{\partial x^2}+\eta u\frac{\partial u}{\partial x}+\alpha\left(u\frac{\partial v}{\partial x}+v\frac{\partial u}{\partial x}\right)=0},\\
        \displaystyle{\frac{\partial v}{\partial t}+\mu\frac{\partial^2 v}{\partial x^2}+\xi v\frac{\partial v}{\partial x}+\beta\left(u\frac{\partial v}{\partial x}+v\frac{\partial u}{\partial x}\right)=0},
    \end{cases}
\end{equation}
where $u\frac{\partial u}{\partial x}$ is the non-linear convection term, and $1/\delta$ and $1/\mu$ are reciprocals of Reynolds numbers. The desired functions to be solved there are velocity components. Our equations \eqref{eq:Burgers} are simpler than (\ref{moregeneralburger}), and can be solved by observing that $E(R,\lambda)\equiv E_1(R,\lambda)+E_2(R,\lambda)$ as well as the solution for $E(R,\lambda)$ being known. 

Solving for individual $E_r(R,\lambda)$ would also require initial condition $E_r(R,0)$ given by the undeformed theory, which would require us to solve the full integral equations. However, in the UV limit \cite{Fendley:1991ve,Fendley:1992dm}, we have 
\begin{equation}
\label{initialconditions}
    E_r(m_rR\rightarrow 0,\lambda = 0) \sim -\frac{1}{\pi R}\bigg[ \mathcal{L}\bigg(\frac{x_r}{1+x_r}\bigg) - \mathcal{L}\bigg(\frac{y_r}{1+y_r}\bigg)\bigg] \equiv - \frac{a_r}{\pi R},
\end{equation}
where $\mathcal{L}(x)$ is the Rogers dilogarithm function
\begin{equation}
    \mathcal{L}(x) \equiv -\frac{1}{2}\int_0^x dy\bigg[\frac{\ln y}{1-y} + \frac{\ln(1-y)}{y}\bigg]
\end{equation}
and $x_r,y_r$ are solutions of some algebraic equations given in \cite{Fendley:1992dm}. 

Therefore, we can find the leading term of $E_r(R,\lambda)$ in the same double scaling limit as in \eqref{fullE(R,lambda)d}:
\begin{equation}\label{perturbative energy}
	E_r(R,\lambda) =
	\frac{3 a_r R}{\pi^2 c\lambda} \left( \sqrt{1 - \frac{2\pi \lambda c}{3R^2}} - 1\right), \quad R\rightarrow 0, \quad \lambda \rightarrow 0, \quad \frac{\lambda}{R^2} \sim O(1).
\end{equation}

We comment on a standard way to somewhat implicitly solve these equations \eqref{eq:Burgers} by finding its characteristic curves which is an integral curve parametrized by a real number $s$ from solving the ordinary \textit{Lagrange-Charpit equations} (or \textit{characteristic equations}):
\begin{equation}
\label{Lagrange-Charpit}
\begin{split}
      \frac{d\lambda(s)}{ds}&=1,    \\
     \frac{dR(s)}{ds}&=-E(R,\lambda),  \\
      \frac{dE_{1,2}(s)}{ds}&=0,
      \end{split}
\end{equation}
where the last equation can be easily seen from the chain rule and \eqref{eq:inviscid}. By demanding the absence of $\lambda(s=0)=0$, the first equation in (\ref{Lagrange-Charpit}) implies $\lambda(s)=s$. The second and the third equations can be solved trivially\footnote{In the following discussion on the method of characteristics, we will suppress arguments of each function [e.g., $E(R,\lambda)\equiv E$] to emphasize that one should treat it as the most na\"ive function. Namely, one is \textit{not} supposed to solve the second equation in \eqref{Lagrange-Charpit} at face value, which is
\begin{equation}
    \frac{dR(s)}{ds}=-E(R,\lambda(s)=s)\simeq\frac{R}{2s}\left(1-\sqrt{1-\frac{2\pi s c}{3R^2}}\right),
\end{equation}
where $\simeq$ is from \eqref{fullE(R,lambda)d}.
}
\begin{equation}
\begin{split}
    R&=Es+C_1,\\
    E_{1,2}&=C_2,
    \end{split}
\end{equation}
and the second expression show that the ground state energies $E_{1,2}$ are invariant along the characteristic curves, which are straight lines with slopes $E$ in the $(s,R)$-plane as shown by the first expression. In \eqref{fullE(R,lambda)}, $E(R,\lambda)$ monotonically increases with $R$ when $R$ is small\footnote{In fact, it also monotonically increases when $R$ is large, as we will show shortly.}, so that the characteristic lines never intersecting each other on the $(s,R)$-plane. As a result, we do not expect phenomena such as wave steepening or shock singularities.

We know from \cite{Fendley:1991ve,Fendley:1992dm} that in the IR limit, for one species of soliton with mass $m$,
\begin{equation}
\label{initialconditionlarge}
    E(R\rightarrow\infty)=-\frac{m}{\pi}\int^{\infty}_{-\infty}d\theta\cosh\theta e^{-R\cosh\theta},
\end{equation}
which is just $-\frac{2m}{\pi}K_1(R)$, in terms of the modified Bessel function of the second kind \cite{gradshteyn2014table}. Given this initial condition at $\lambda=0$, the inviscid Burgers' equation \eqref{eq:inviscid} has the general solution $E(R,\lambda)$ \textit{algebraically} solving this functional equation:
\begin{equation}
    E(R,\lambda)=-\frac{2m}{\pi}K_1(R-\lambda E),
\end{equation}
which is unfortunately transcendental. However, 
\begin{equation} \partial_{\lambda}E(R,\lambda)=\frac{m}{\pi}\left[K_0(R-\lambda E)+K_2(R-\lambda E)\right]E
\end{equation}
is always positive due to $E(R,\lambda)>0$, and from \eqref{eq:inviscid} we know that $\partial_R E(R,\lambda)>0$, when $R\rightarrow\infty$. Overall, since $E(R,\lambda)$ monotonically increases with $R$ when $R$ is both small and large, it is reasonable to assume that this monotonicity holds for all $R$.

The general solutions to \eqref{Lagrange-Charpit}, when $R$ is not extremal, may be found in terms of $R$ and $s$ by noting that $C_2$ must be an arbitrary function of $C_1$, namely $C_2(C_1)$. We have $C_1 = R-Es$ and $E_{1,2}=C_2(R-Es)$. Now we need to determine the function form from an initial condition $E_{1,2}(R(s=0),0)$.

Unfortunately, \eqref{initialconditions} [or \eqref{initialconditionlarge}] is just an initial condition at a single point, instead of telling us the complete $R$-profile of $E_{1,2}(R,0)$ (analogous to the spatial distribution of temperature at time $t=0$ for the heat equation.) This $R$-profile is also beyond the perturbative treatment around $R=0$ described previously \eqref{perturbative energy}. Hence, the method of characteristics requires additional data than we are able to present here analytically. However, if one manages to numerically obtain the profile $E_r(R,0)$ form the full integral equations using techniques discussed in \cite{Klassen:1989ui, Zamolodchikov:1989cf, zamolodchikov1991thermodynamic, Fendley:1991xn, Dorey:1996re}, then it is possible to obtain the complete solutions $E_r(R,\lambda)$. Finally, we again note that the above method easily generalizes to $n>2$.

\subsection{Connections to \texorpdfstring{$\mathcal{N} = (1,1)$}{} models}
In this subsection, we briefly extend the $\mathcal{N} = (2, 2)$ TBA formalism from the previous subsections to two-dimensional $\mathcal{N} = (1, 1)$ integrable models \cite{Moriconi:1995aj}. Melzer \cite{Melzer:1994qp} rigorously showed that the $\mathcal{N} = (2, 2)$ integrable systems derived by Fendley and Intriligator \cite{Fendley:1991ve} are related to $\mathcal{N} = (1, 1)$ integrable systems via a ``folding'' procedure. The folding procedure relates certain $\mathcal{N} = (2,2)$ TBA systems with $2n$ types of particles to $\mathcal{N} = (1,1)$ TBA systems with $n$ types of particles.  

This folding procedure requires the TBA system to possess the following symmetries:
\begin{equation}
  m_{a}=m_{2 n+1-a}, \quad  \phi_{a, b} (\theta) = \phi_{2 n+1-a, 2 n+1-b} (\theta),\quad a,b = 1,\cdots,2n.
\end{equation}

Folding this TBA system would mean that we have half the number of particles $a =1, \ldots, n$ and the folded kernel is
\begin{equation}
  \phi_{a, b}^{\mathrm{folded}} (\theta) =  \phi_{a, b} (\theta) +  \phi_{a, 2 n+1-b} (\theta), \quad a, b=1,2, \ldots, n.
\end{equation}

Inspired from this, we study the $\mathcal{N} = (1,1)$ theory from folding the LG models with superpotential
\begin{equation}
    W(X) = \frac{X^{2k}}{2k} - \beta X.
\end{equation}
In this subsection, we will achieve the following:
\begin{enumerate}
    \item The ground state energy's flow equation obeys the inviscid Burgers’ equation.
    \item The folding of the $T\oT$ deformed theory is the same as the $T\oT$ deformation of the folded theory. 
\end{enumerate}

First, we construct the TBA system for the folded $T\oT$ deformed theory and derive the ground state energy's flow equation. 

For the undeformed theory, the TBA system contains $2k-2$ massive particles with masses
\begin{equation}
    m_a = \sin\bigg(\frac{r\pi}{2k-1}\bigg) M, \,\,\, r = 1,\cdots, 2k-2,
\end{equation}
and two massless particles labeled by $0,\overline{0}$. Identifying $\overline{0}$ with $2k-1$, then clearly we have
\begin{equation}
    m_a = m_{2k-1-a}.
\end{equation}

It it straightforward to check $\phi_{a,b} = \phi_{2k-1-a,2k-1-b}$ using the $\mathcal{N} = (2,2)$ kernels (\ref{ker1}) and (\ref{ker2}).

With the $T\oT$ deformation, the masses remain unchanged while\footnote{Notice that since $m_0 = m_{2k-1} = 0$, $\phi_{r,l}$ remains unchanged which is consistent with the previous notation.}
\begin{equation}
    \phi_{a,b}(\theta,\lambda) = \phi_{a,b}(\theta,0) + \lambda m_a m_b \cosh(\theta), \quad a,b = 0,\cdots,2k-1.
\end{equation}
From the symmetry of the masses $m_a = m_{2k-1-a}$, we learn that the symmetry of the kernels also holds under the $T\oT$ deformation:
\begin{equation}
\begin{aligned}
    \phi_{r,s}(\theta,\lambda) &= \phi_{r,s}(\theta,0) + \lambda m_r m_s \cosh(\theta) \\
    &= \phi_{2k-1-a,2k-1-b}(\theta,0) + \lambda m_{2k-1-a} m_{2k-1-b} \cosh\theta \\
    &= \phi_{2k-1-r,2k-1-s}(\theta,\lambda).
\end{aligned}
\end{equation}

The folding of the $T\overline{T}$ deformed theory has a TBA system containing $k$ particles with mass spectrum
\begin{equation}
    m_a^{\text{folded}} = M \sin\bigg(\frac{a\pi}{2k-1}\bigg), \,\,\, a= 0,1,\cdots,k-1,
\end{equation}
together with the kernels
\begin{equation}\label{N=2toN=1}
\begin{aligned}
    \phi^{\text{folded}}_{a,b}(\theta,\lambda) &= \phi_{a,b}(\theta,\lambda) + \phi_{a,2k-1-b}(\theta,\lambda) \\
    &= \phi_{a,b}(\theta,0) + \phi_{a,2k-1-b}(\theta,0) + \lambda m_a (m_b+m_{2k-1-b})\cosh\theta \\
    &= \phi_{a,b}^{\text{folded}}(\theta,0) + 2\lambda m_a m_b \cosh\theta, \quad a,b = 0,\cdots,k-1.
\end{aligned}
\end{equation}
It is then straightforward to show that the $T\oT$ deformation shifts the radius $R$ in the integral equation by an energy-dependent term $\lambda E(R)$ as in (\ref{N=2TBA}). Therefore, the flow equation of the ground state energy obeys the usual inviscid Burgers’ equation. 
 
Next, we want to show that the folding of a $T\oT$ deformed theory is indeed the same as $T\oT$ deformation of the folded theory. By definition, the $T\oT$ deformation is given in terms of the S-matrix instead of a kernel. A priori, we do not know what the $T\oT$ deformation of the kernel $\phi_{a,b}$ should be without knowledge of the S-matrix. We will use the relations in \cite{Moriconi:1995aj, Moriconi:1996tn} between the kernels and S-matrix to show this is indeed the case. 
 
The undeformed $\mathcal{N} = (1,1)$ theory's S-matrix is 
\begin{equation}
    S^{[ij]}(\theta) = S_{BF}^{[ij]}(\theta)S_B^{[ij]}(\theta)
\end{equation}
where $[ij]$ tell us which supermultiplets the solitons in the scattering belong to, $S_B^{[ij]}$ is a bosonic S-matrix and $S_{BF}^{[ij]}$ is a piece of the S-matrix that mixes bosons and fermions. The solitons' masses are
\begin{equation}
    m_a^{\mathcal{N}=(1,1)} = \frac{\sin(a\pi/(2n+1))}{\sin(\pi/(2n+1))}, \,\,\, a = 1,\cdots n.
\end{equation}

The deformation is given by multiplying the S-matrix by 
\begin{equation}
    S^{[ij]}(\theta,\lambda) = S^{[ij]}(\theta,0) e^{i\lambda m_i m_j \cosh\theta},
\end{equation}
and the extra phase factor can be absorbed into the bosonic S-matrix 
\begin{equation}
    S^{[ij]}_B(\theta,\lambda) = S_B^{[ij]}(\theta,0) e^{i\lambda m_i m_j \cosh\theta}.
\end{equation}
Then we use the relation between the kernel and the S-matrix:
\begin{equation}
\phi_{ab}^{\mathcal{N}=1}(\theta,\lambda) = \frac{\partial}{\partial \theta} \text{Im} \ln \bigg(\frac{S_B^{[ij]}(\theta,\lambda)g^{[ij]}(\theta)}{\sinh\theta}\bigg), \quad a = 1,\cdots,n,
\end{equation}
where $g^{[ij]}(\theta)$ is some integral expression inside $S_{BF}^{[ij]}(\theta)$ and thus independent of $\theta$. In addition, there will be a single auxiliary massless particle labelled by $0$ whose kernel $\phi^{\mathcal{N}=1}_{a,0}(\theta)$ is independent of $S_{B}^{[ij]}(\theta,\lambda)$ and we conclude that the $T\oT$ deformation of the folded theory is the same as the folding of the $T\oT$ deformed theory up to a re-scaling in $\lambda$.

\subsection{Deformed supersymmetric indices}
\label{sec:index}
To conclude this section, we will study various supersymmetric indices under the $T\oT$ deformation for LG models. 

The most famous supersymmetric index is the Witten index \cite{Witten:1982df}. The Witten index is invariant under the $T\oT$ deformation because the deformation does not lift the energy degeneracy between bosons and fermions (although the energy itself does flow), which implies the structure of the ground states remain unchanged. 

There are other indices that are interesting to study under the deformation, such as the CFIV index \cite{Cecotti:1992qh} and the elliptic genus \cite{Schellekens:1986yi, Schellekens:1986yj, Schellekens:1986xh, Witten:1986bf}. We will consider these two indices in the following subsections. 

\subsubsection{CFIV index}
First, consider 
\begin{equation}
\label{eq:index}
    Z(\alpha,\beta) = \text{Tr} e^{i\alpha F}e^{-\beta H}.
\end{equation}
For $\alpha = \pi$, this is simply the Witten index
\begin{equation}
    I_0 \equiv Z(\pi,\beta) = \text{Tr} (-1)^F e^{-\beta H}.
\end{equation}
Taking a derivative of \eqref{eq:index} with respect to $i\alpha$ and setting $\alpha = \pi$, we arrive at the CFIV index, which is invariant under $D$-term perturbations introduced in \cite{Cecotti:1992qh},
\begin{equation}
\label{eq:I1}
    I_1(\beta) = \text{Tr} (-1)^F F e^{-\beta H}.
\end{equation}

One can consider generalization of $I_1 (\beta)$ by taking more derivatives, 
\begin{equation}
    I_l(\beta) = \text{Tr} (-1)^F F^l e^{-\beta H}, \quad l \geq 2,
\end{equation}
however, as shown in \cite{Cecotti:1992qh}, these quantities are not invariant under $D$-term perturbations and are not indices. 

For a theory with a mass gap, sometimes there is a vacuum degeneracy. Then we can consider soliton configurations interpolating between the vacuum labelled by $a$ on the left and the vacuum labelled by $b$ on the right. Thus, we can define $I_l$ for each pairwise soliton configuration
\begin{equation}
	(I_l)_{ab} = \text{Tr}_{ab}(-1)^F F^l e^{-\beta H}
\end{equation}
so that we now have $n\times n$ ($n$ is the number of vacua) matrices rather than a number. 

For the Witten index, $(I_0)_{ab}$ is just a diagonal matrix. This is exactly because for $a\neq b$, a BPS soliton is required to connect the two different vacua. However, as we know, this is a two-dimensional representation of $\mathcal{N} = (2,2)$ supersymmetry with non-zero energy and will not contribute to the Witten index. Furthermore, since in our case there is a $\mathbb{Z}_n$ symmetry that relates the $n$ vacua, $(I_0)_{ab}$ is proportional to the identity matrix.

For the CFIV index, we obtain a general $n\times n$ matrix and are interested in their eigenvalues. In practice, we keep track of the eigenvalues by introducing a weight for each soliton type. For instance, if the soliton is charged under some topological symmetry with charge $\mathcal{T}$, then we can introduce a chemical potential $e^{i\Theta \mathcal{T}}$, expand the final answer in terms of $e^{i m \Theta}$ with coefficients being the eigenvalues of $(I_1)_{ab}$. For this model, we consider the superpotential $W(X,\beta) = \frac{X^{n+1}}{n+1} - \beta X$ which leads to $n$ vacua at $X = e^{2\pi i j/n} \beta^{1/n}, j = 1\cdots n$. As usual, we call a soliton type-$r$ if the two vacua are related to each other by satisfying $X_a = e^{2\pi i r/n}X_b$. The $p$-th eigenvalues of $(I_1)_{ab}$ are found by introducing a weight $e^{2\pi i r p N_r/n}$, where $N_r$ are the number of type-$r$ solitons.

To make connections to integrable theories, we consider the free energy $\mathcal{F}_{\mu_a}(\beta)$ with chemical potential $\mu_a$:
\begin{equation}
    - \beta \mathcal{F}_{\mu_a}(\beta) = \text{ln} \text{Tr}(e^{\beta\sum_a \mu_a N_a} e^{-\beta H}).
\end{equation}

The exact expression $\beta \mathcal{F}_{\mu_a}(\beta)$ can be computed via TBA using the exact S-matrix

\begin{equation}
    \ln \text{Tr}(e^{\beta \sum_a \mu_a F_a} e^{-\beta H}) = -\sum_a m_a L \int \frac{d\theta}{2\pi}\cosh\theta \ln(1+e^{\beta \mu_a - \epsilon_a(\theta)}),
\end{equation}
where the $\epsilon_a(\theta)$ are the solutions to the coupled integral equations:
\begin{equation}
    \epsilon_a(\theta) = m_a \beta \cosh(\theta) - \sum_b \int \frac{d\theta'}{2\pi} \phi_{ab}(\theta - \theta') \ln(1+e^{\beta\mu_b - \epsilon_b(\theta')}).
\end{equation}

Recall from previous subsections that the ground state's momentum vanishes
\begin{equation}
    - \sum_a m_a L \int \frac{d\theta}{2\pi} \sinh(\theta) \ln(1+e^{\beta\mu_a - \epsilon(\theta)}) = 0.
\end{equation}
In the vanishing chemical potential case ($\mu_a = 0$), we argue that this integral vanishes because of  $\epsilon_a(\theta) = \epsilon_a(-\theta)$. The same occurrence with vanishing momentum happens here since the chemical potential only favors particular species of particles but not the rapidity. Thus the $\theta \rightarrow -\theta$ parity symmetry is unaffected by the chemical potential. For a vanishing chemical potential, we similarly derive a flow equation of $-\beta \mathcal{F}_{\mu_a}(\beta)$ by introducing the quantity
\begin{equation}
    F_p(\beta,\lambda,\alpha) = - \sum_a m_a \beta \int \frac{d\theta}{2\pi} \cosh(\theta) \ln(1+e^{\beta\mu_a-\epsilon_a(\theta)}) = \frac{\beta}{L} \ln \text{Tr}(e^{i\alpha F + 2\pi i r p N_r/N}e^{-\beta H}),
\end{equation}
where in the last step we have chosen $\mu_a$ such that $\beta \sum_a \mu_a N_a = i\alpha F + 2\pi i r p N_r/N$, and $p=0,\dots,n-1$. Then we find the flow equation for $F_p(\beta,\lambda,\alpha)$:
\begin{equation}
    \partial_\lambda F_p(\beta,\lambda,\alpha) + F_p(\beta,\lambda,\alpha) \partial_\beta F_p(\beta,\lambda,\alpha) = 0.
\end{equation}
Unlike the Witten index, when being placed on $S^1$ with radius $L$, $I_1$ from \eqref{eq:I1} does scale linearly with $L$. It is useful to introduce another quantity  
\begin{equation}
    Q_p(\beta,\lambda) = \partial_\alpha F_p(\beta,\lambda,\alpha)_{\alpha = \pi} = \frac{\beta}{L} \partial_\alpha \ln \text{Tr}(e^{i\alpha F + 2\pi i r p N_r/N - \beta H})_{\alpha = \pi},
\end{equation}
which is clear in the thermodynamic limit $L\rightarrow \infty$. Taking a derivative of the flow equation with respect to $\alpha$ evaluated at $\alpha = \pi$, we find
\begin{equation}
    \partial_\lambda Q_p(\beta,\lambda) + Q_p(\beta,\lambda) \partial_\beta \bigg(\frac{\beta}{L} \ln I_0\bigg) + \frac{\beta}{L} (\ln I_0) \partial_\beta Q_p(\beta,\lambda) = 0.
\end{equation}
After being simplified, the flow equation for $Q_p (\beta, \lambda)$ is
\begin{equation}
    \partial_\lambda Q_p(\beta,\lambda) + \frac{1}{L} (\ln I_0) Q_p(\beta,\lambda) + \frac{\beta}{L}(\ln I_0) \partial_\beta Q_p(\beta,\lambda) = 0,
\end{equation}
where we have the same $I_0$ for all $p$ since $(I_0)_{ab}$ is proportional to the identity and can be simultaneously diagonalized with $(I_1)_{ab}$ with $N$ identical eigenvalues.\footnote{We will denote the eigenvalues as $I_0$ and we have used the fact that $I_0$ is independent of $\beta$ and $\lambda$.} In the thermodynamic limit $L\rightarrow \infty$, we find
\begin{equation}
    \partial_\lambda Q_p(\beta,\lambda) = 0.
\end{equation}
In other words, the CFIV index does not flow under $T\overline{T}$ deformation and, perhaps, is not a surprise after all. As argued in \cite{Cecotti:1992qh}, the CFIV index is dependent of $F$-term perturbations and is independent of $D$-term perturbations. These two facts combined imply that the CFIV index will not flow under $T\overline{T}$ deformation.

Now, let us consider $I_l$ for $l\geq 2$. These are not invariant under $D$-term perturbations making them not indices. Therefore, it is natural to expect these $I_l$'s would flow under the $T\overline{T}$ deformation. Indeed, TBA allows one to derive a set of recursive flow equations for these $I_l$'s. For instance, consider $I_2$ and define 
\begin{equation}
    Q_{2,p} = \partial_\alpha^2 F_p(\beta,\lambda,\alpha)|_{\alpha = \pi} = \frac{\beta}{L}\bigg(\frac{I_{1,p}^2}{I_0^2} - \frac{I_{2,p}}{I_0}\bigg).
\end{equation}

Taking two derivatives with respect to $\alpha$ on both sides of the flow equation for $F_p(\beta,\lambda,\alpha)$, we obtain
\begin{equation}
    \partial_\lambda Q_{2,p} + (Q_{2,p} + \beta \partial_\beta Q_{2,p}) \frac{\ln I_0}{L} + 2 Q_{1,p} \partial_\beta Q_{1,p}  = 0.
\end{equation}

If $I_{2,p}$ scales linearly with $L$ in the thermodynamic limit $L\rightarrow\infty$, then $\frac{Q_{2,p}}{L} \rightarrow \beta\frac{I_{1,p}^2}{L^2} = - \frac{1}{\beta} Q_{1,p}^2$. As a result, we have
\begin{equation}
    \beta \partial_\lambda (I_2/L) - 2I_0(1 - \ln I_0) Q_{1,p} \partial_\beta Q_{1,p} = 0,
\end{equation}
where we used $\partial_\lambda I_0 = \partial_\lambda I_{1,p} = 0$.

\subsubsection{Elliptic genus}
Another interesting index to study in $\mathcal{N} = (2,2)$ theory is the elliptic genus \footnote{The elliptic genus was first shown not to flow under the $T\oT$ deformation in \cite{Datta:2018thy} by showing the existence of a BPS-like sector in the spectrum. In our setting, we use the TBA as an alternative proof to further support the elliptic genus not flowing under the $T\oT$ deformation.} defined by
\begin{equation}
    \text{Tr} e^{i\alpha_L F_L}(-1)^{F_R} e^{-\beta H}.
\end{equation}
However, to study this theory, we must abandon the LG model with superpotential
\begin{equation}
    W(X) = \frac{X^{n+1}}{n+1} - \beta X
\end{equation}
as the elliptic genus is not well-defined for gapped theories \cite{Fendley:1993pi}. 

Instead, we consider the LG model with superpotential
\begin{equation}
    W = g X^{k+2}
\end{equation}
where $g$ is the coupling constant. These theories are conjectured to be integrable supported with highly non-trivial checks in \cite{Fendley:1993pi}.

All excitations are massless with $H=|P|$ in the system. The left-movers are in the doublet representation $(u_L(\theta),d_L(\theta))$ of the left-moving $\mathcal{N}=(0, 2)$ supersymmetry with $H = -P = Me^{-\theta}$. The right-movers are in the same representation of the right-moving $\mathcal{N}=(0, 2)$ supersymmetry, but with $H = P = Me^{\theta}$. The S-matrix contains three parts: $S_{LL}$ ($S_{RR}$) which encodes the scattering among the left(right)-movers, and $S_{LR}$ which encodes the scattering between left-movers and right-movers. The exact form of the S-matrix is irrelevant for this discussion and can be found in \cite{Fendley:1993pi}. When the excitations are massless, as pointed out in \cite{Cavaglia:2016oda}, the $T\overline{T}$ deformation manifests itself as
\begin{equation}
    S_{ij}^{kl}(\theta, \lambda) = S_{ij}^{kl}(\theta, 0) e^{i\delta^{(\lambda)}_{ij}(\theta)}
\end{equation}
and
\begin{equation}
    \delta_{ij}^{(\lambda)}(\theta_i - \theta_j) = - 2\lambda p_i^{(+)} p_j^{(-)},
\end{equation}
where $p_i^{(+)}$ and $p_j^{(-)}$ are the momenta of right- and left-moving particles. Thus, we see only $S_{LR}$ is modified under $T\overline{T}$ deformation with
\begin{equation}
    \delta_{LR}^{(\lambda)}(\theta_L - \theta_R) = -2\lambda M^2 e^{\theta_L - \theta_R}.
\end{equation}

For simplicity, we start with the simplest case with $k=1$:
\begin{equation}
    W = g X^3.
\end{equation}

The undeformed integral equations are
\begin{equation}
\label{eq:integraleqgenus}
    \epsilon_a(\theta) = \nu_a(\theta) - \sum_{b} l_{ab} \int \frac{d\theta'}{2\pi} \frac{1}{\cosh(\theta-\theta')} \ln(1+\lambda_a e^{-\epsilon_a(\theta')}),
\end{equation}
where $\nu_a(\theta)$'s, $\lambda_a$'s, $l_{ab}$ and the index $a$ are encoded in the following diagram:
\begin{equation}
\label{eq:diagram}
\begin{tikzpicture}
\draw[black, thick] (-0.6,0) -- (0.6,0);
\draw (-0.8,0) node {L};
\draw (0.8,0) node {R};
\draw[black, thick] (-1.0,0.2) -- (-1.6,0.8);
\draw[black, thick] (-1.0,-0.2) -- (-1.6,-0.8);
\draw[black, thick] (1.0,0.2) -- (1.6,0.8);
\draw[black, thick] (1.0,-0.2) -- (1.6,-0.8);
\draw (-1.8,1.0) node[circle, draw, thick] {};
\draw (-2.6,1.0) node {$e^{i\alpha_L}$};
\draw (1.8,1.0) node[circle, draw, thick] {};
\draw (2.6,1.0) node {$e^{i\alpha_R}$};
\draw (-1.8,-1.0) node[circle, draw, thick] {};
\draw (-2.6,-1.0) node {$e^{-i\alpha_L}$};
\draw (1.8,-1.0) node[circle, draw, thick] {};
\draw (2.6,-1.0) node {$e^{-i\alpha_R}$};
\end{tikzpicture}.
\end{equation}
The index $a$ runs over each node:
\begin{equation}
    \nu_a(\theta) = \begin{cases}
    0 & \text{if node $a$ is open}, \\
    \frac{1}{2}M\beta e^{-\theta} & a = L, \\
    \frac{1}{2}M\beta e^{\theta} & a = R.
    \end{cases}
\end{equation}
The $\lambda_a$'s are given by the phases next to the four open nodes, and are equal to $1$ for $a=L,R$. Here $l_{ab} = 1$ if two nodes are connected and is $0$ otherwise. The TBA system is used to compute the following quantity:
\begin{equation}
\begin{aligned}
    c(\alpha_L,\alpha_R;M\beta) &\equiv \frac{6\beta}{\pi L} \log \text{Tr}e^{i\alpha_L F_L} e^{i\alpha_R F_R} e^{-\beta H} \\
    &= \frac{3}{\pi^2} \sum_a \int d\theta \nu_a(\theta) \log(1+\lambda_a e^{-\epsilon_a(\theta)}).
\end{aligned}
\end{equation}

The coupling between the $L$ and $R$ nodes is a result of the S-matrix $S_{LR}$ and with the $T\overline{T}$ deformation, we expect the undeformed kernel $\phi_{LR} = \frac{1}{\cosh(\theta-\theta')}$ to be modified by 
\begin{equation}
    \phi_{LR} \rightarrow \phi_{LR} - 2\lambda M^2 e^{\theta - \theta'}.
\end{equation}

Deriving a flow equation at any $\alpha_{L,R}$ for $c(\alpha_L,\alpha_R;M\beta)$ is arduous due to not being invariant under supersymmetry-preserving deformations. However, for the special value $\alpha_R = \pi$, $c(\alpha_L,\pi;M\beta)$ is the elliptic genus which is invariant under supersymmetry-preserving deformations. Following this logic, we will show $c(\alpha_L,\pi;M\beta)$ does not flow under the $T\overline{T}$ deformation.

Let us first fix $\alpha_R = \pi$. For the undeformed TBA, with $\lambda = 0$, the theory has a special solution \cite{Fendley:1993pi}
\begin{equation}
    e^{-\epsilon_R(\theta)} = \epsilon_{\pm F_R}(\theta) = 0,
\end{equation}
where by $\pm F_R$, we mean the two nodes to the right of the node $R$ in \eqref{eq:diagram}. To see this, in the integral equations \eqref{eq:integraleqgenus} of $\epsilon_{\pm F_R}(\theta)$, we have
\begin{equation}
    \epsilon_{\pm F_R}(\theta) = - \int \frac{d\theta'}{2\pi} \frac{1}{\cosh(\theta-\theta')} \ln(1+e^{-\epsilon_R(\theta)}) = 0
\end{equation}
while for $e^{-\epsilon_R(\theta)}$, we find
\begin{equation}
\begin{aligned}
    e^{-\epsilon_R(\theta)} &= \exp\bigg[\int \frac{d\theta'}{2\pi} \frac{1}{\cosh(\theta-\theta')} \ln(2+2\cos\alpha_R) + \text{finite}\bigg] \\
    & = \sqrt{2+2\cos\alpha_R} \times (\text{finite}) = 0 \,\,\, \text{for $\alpha_R = \pi$}.
\end{aligned}
\end{equation}

In this solution, the right-movers completely decouple from the left-movers as $\ln(1+e^{-\epsilon_R(\theta)}) = 0$. In other words, the terms related to $\epsilon_R(\theta)$ vanishes in the integral equation \eqref{eq:integraleqgenus} of $\epsilon_L(\theta)$. It is straightforward to check that this remains true under the $T\overline{T}$ deformation. This is again not a surprise. What the $T\overline{T}$ deformation does is simply to introduce a non-trivial coupling between left-movers and right-movers. If the two sectors are already decoupled in the TBA system for the elliptic genus, then $T\overline{T}$ deformation simply will not have any effects on the TBA system. 

It is not difficult to see that this argument generalizes to the superpotential $W = g X^{k+2}$ for $k \geq 2$. The only difference in the TBA system is now there are extra $k-1$ open nodes linearly connecting nodes $L$ and $R$. However, since $\alpha_R = \pi$, the right-moving sector would decouple anyways so the $T\overline{T}$ deformation will not change the TBA system for the remaining left-moving sector. 

\section{Conclusion and future directions}
\label{sec:conclusion}
In this paper, we have further explored several aspects of $T\oT$ deformed SCFTs, integrable supersymmetric models and their indices. We have calculated the deformed two- and three-point correlators for two-dimensional Euclidean $\mathcal{N} = (0, 2)$ SCFTs in the spirit of \cite{Kraus:2018xrn}. However, unlike in \cite{Kraus:2018xrn}, we did not look at the bulk AdS$_3$ supergravity side \cite{Achucarro:1987vz, Kraus:2007vu} as an alternative method to obtain the deformed correlators. Additionally, studying how the correlators change under different solvable irrelevant deformations, such as using \cite{Jiang:2019trm}'s construction for the deformed multiplets in $J\oT$, $\bar{J}T$ and \cite{He:2019vzf}'s systematic methods to evaluate $J\oT, \bar{J}T$ correlators via conformal perturbation theory, seems tractable. Also, it would be interesting to non-perturbatively calculate the deformed SCFT correlators \'{a} la Cardy \cite{Cardy:2019qao}.

Our perturbative analysis of the $\mathcal{N} = (2,2)$ $\mS$-multiplets near the superconformal fixed point does not completely explain the full feature of the central current generated under the $T\oT$ deformation and does not allow one to convincingly determine the fate of the chiral ring and twisted ring in the deformed theory. We believe a non-perturbative study of the $T\oT$ deformation is required for further exploration. 

We also studied the $T\oT$ deformed $\mathcal{N}= (1,1)$ and $\mathcal{N}=(2,2)$ supersymmetric integrable models' S-matrices, ground state energies and common indices via TBA. We derived a flow equation for the deformed ground state energy and showed that several $D$-term independent indices do not flow. Among these indices, the CFIV index is special as it is not a topological index but can be derived solely from the topological data (topological-anti-topological fusions), i.e., the deformed $\mathcal{N} = (2,2)$ chiral ring, via the deformed $tt^*$ equations. It would be interesting to study more aspects of the $T\oT$ deformed (twisted) chiral ring in relation to TQFTs \cite{Witten:1988xj, Witten:1991zz, Santilli:2018xux, Santilli:2020qvd}. 

While we have analyzed a class of deformed $\mathcal{N} = (2,2)$ two-dimensional integrable models via TBA or ABA, there are other interesting models one can consider such as the following: $\mathbb{Z}_n$ generalizations of the supersymmetric sine-Gordon model, supersymmetric $\mathbb{C}\mathbb{P}^{n-1}$ sigma models and $SU(2)_{l} \otimes SU(2)_{k} / SU(2)_{k+l}$ coset models. Also, it would be interesting to use TBA to numerically study excited states in these models. Numerical solutions for the excited states of the TBA equations have been studied in \cite{Dorey:1996re}, which is extended to supersymmetric integrable models in \cite{Fendley:1997ys}.
 
We studied the $T\oT$ deformed S-matrices for a certain class of $\mathcal{N} = (1,1)$ and $\mathcal{N} = (2,2)$ two-dimensional integrable models, and a natural question is to consider these same theories when there is a boundary present to determine, with suitable boundary conditions, the boundary reflection matrices on how much integrability and supersymmetry are preserved. In the case for two-dimensional $\mathcal{N} = (1,1)$ integrable models \cite{Moriconi:1996tn}, reminiscent of the S-matrix, the R-matrix is 
\begin{equation}
    R^{[ij]}(\theta) = R^{[ij]}_{BF} (\theta) R^{[ij]}_B (\theta),
\end{equation}
where $R^{[ij]}_B(\theta)$ is the reflection matrix for the bosonic part and $R^{[ij]}_{BF}(\theta)$ describes the relative amplitudes for bosons and fermions when scattering off the boundary. The deformed R-matrix obeys the usual unitarity condition, boundary Yang-Baxter equations and crossing symmetry in terms of the deformed S-matrix. It would be interesting to study more on boundary supersymmetric integrable models in the context of \cite{Fendley:1990zj, Fendley:1993zt, Fendley:1994cz, Warner:1995ay, Inami:1995np, Moriconi:1996tn} under the $T\oT$ deformation. 

We hope to return to these open problems in future works. 

\acknowledgments
We are deeply grateful to Christian Ferko for a careful read of the manuscript, sharing unpublished notes and providing insightful comments. We thank Thomas T. Dumitrescu, Alexander Frenkel, Ken Intriligator, Per Kraus, John McGreevy and Sridip Pal for useful discussions. S.E. acknowledges funding from the Bhaumik Institute and thanks UCSD for hospitality where part of this work was completed. H.-Y.S. is supported by the the Simons Collaborations on Ultra-Quantum Matter, grant \#651440 from the Simons Foundation. Z.S. acknowledges the support from the US Department of Energy (DOE) under cooperative research agreement DE-SC0009919 and Simons Foundation award  \#568420. We also appreciate the organizers and participants of the workshop on ``\emph{$T\oT$ Deformation and Integrability}'' for enlightening talks at APCTP during November 8 - 14, 2020.

\newpage
\appendix 

\section{Commutation relations for two-dimensional \texorpdfstring{$\mathcal{N} = (2,2)$}{} SCFT}
In this appendix, we review how one arrives at the commutation relations for two-dimensional $\mathcal{N} = (2,2)$ SCFT in \cite{Dumitrescu:2011iu}. For any superfield $S$
\begin{equation}
\label{MasterEq}
    [\xi^{+} Q_{+}+\xi^{-} Q_{-}-\overline{\xi}^{+} \overline{Q}_{+}-\overline{\xi}^{-} \overline{Q}_{-}, S]=i\left(\xi^{+} \mathcal{Q}_{+}+\xi^{-} \mathcal{Q}_{-}-\overline{\xi}^{+} \overline{\mathcal{Q}}_{+}-\overline{\xi}^{-} \overline{\mathcal{Q}}_{-}\right) S,
\end{equation}
where $\mathcal{Q}_{\pm}$ are differential operators and ${Q}_\pm$ are supercharges 
\begin{equation}
\begin{aligned}
\mathcal{Q}_\pm &= \frac{\partial}{\partial \theta^\pm} + \frac{i}{2} \overline{\theta}^\pm \partial_{\pm \pm}, \\ 
\overline{\mathcal{Q}}_\pm &= - \frac{\partial}{\partial \overline{\theta}^\pm} - \frac{i}{2} \theta^\pm \partial_{\pm \pm}. 
\end{aligned}
\end{equation}

For simplicity, to see how \eqref{MasterEq} is used, we will first look at the $\mathcal{N}= (0,2)$ $\mathcal{S}$-multiplet. This amounts to set $\xi^- = \overline{\xi}^- = 0$. The $\mathcal{N} = (0,2)$ $\mathcal{S}$-multiplet contains two real superfields $\mathcal{S}_{++}$ and $\mathcal{T}_{----}$ as well as a complex superfield $\mathcal{W}_-$ which obey the following constraints
\begin{equation}
    \begin{split}
    \partial_{--} \mathcal{S}_{++}&=D_{+} \mathcal{W}_{-}-\overline{D}_{+} \overline{\mathcal{W}}_{-}, \\ 
    \overline{D}_{+} \mathcal{W}_{-}&=C,\\
    \overline{D}_{+} \mathcal{T}_{-----}&=\frac{1}{2} \partial_{--} \mathcal{W}_{-}.
    \end{split}
\end{equation}

Solving the above constraints in terms of components yields
\begin{equation}
    \begin{split}
    \mathcal{S}_{++}&=j_{++}-i \theta^{+} S_{+++}-i \overline{\theta}^{+} \overline{S}_{+++}-\theta^{+} \overline{\theta}^{+} T_{++++}, \\ \mathcal{W}_{-}&=-\overline{S}_{+--}-i \theta^{+}\left(T_{++--}+\frac{i}{2} \partial_{--} j_{++}\right)-\overline{\theta}^{+} C+\frac{i}{2} \theta^{+} \overline{\theta}^{+} \partial_{++} \overline{S}_{+--}, \\ \mathcal{T}_{----}&=T_{----}-\frac{1}{2} \theta^{+} \partial_{--}S_{+--}+\frac{1}{2} \overline{\theta}^{+} \partial_{--} \overline{S}_{+--}+\frac{1}{4} \theta^{+} \overline{\theta}^{+} \partial_{--}^{2} j_{++}.
    \end{split}
\end{equation}
Starting with $\overline{Q}_+$:
\begin{equation}
    \begin{split}
    \lbrack \overline{Q}_+, \mathcal{S}_{++} \rbrack &= i \overline{\mathcal{Q}}_+ \mathcal{S}_{++}, 
    \end{split}
\end{equation}
\begin{equation}
    \begin{aligned}
 \lbrack \overline{Q}_+, j_{++}-i \theta^{+} &S_{+++}-i \overline{\theta}^{+} \overline{S}_{+++}-\theta^{+} \overline{\theta}^{+} T_{++++} \rbrack\\ &= -i\left(\frac{\partial}{\partial \overline{\theta}^+} + \frac{i}{2} \theta^+ \partial_{+ +} \right) \left( j_{++}-i \theta^{+} S_{+++}-i \overline{\theta}^{+} \overline{S}_{+++}-\theta^{+} \overline{\theta}^{+} T_{++++} \right) \\ &= - \overline{S}_{+++} + i\theta^+ T_{++++} + \frac{1}{2} \theta^+ \partial_{++} j_{++} - \frac{i}{2}\overline{\theta}^+\theta^+ \partial_{++} \overline{S}_{+++}.
    \end{aligned}
\end{equation}
So, we arrive at
\begin{equation}
    \begin{split}
 \left[ \overline{Q}_+, j_{++ }\right] &= - \overline{S}_{+++} ,
 \\
\{ \overline{Q}_+, S_{+++ } \} &= - T_{++++} - \frac{i}{2} \partial_{++} j_{++}, 
\\
\{\overline{Q}_+, \overline{S}_{+++} \} &= 0,
\\
\left[ \overline{Q}_+, T_{++++} \right] &= \frac{i}{2} \partial_{++} \overline{S}_{++++}.  
    \end{split}
\end{equation}
The rest of the commutation and anti-commutation relations for the $\mathcal{N} = (0,2)$ superconformal algebra are easily obtainable from the same method. 

Using \eqref{MasterEq}, we tabulate all the commutation and anti-commutations relations for the $\mathcal{N} = (2, 2)$ superconformal algebra. 

For $Q_+$:
\begin{equation}
    \begin{aligned}
            \left[Q_+, j_{++} \right] &= S_{+++}, \\
            \left[Q_+, j_{--} \right] &= S_{+--} + i 2\sqrt{2} \overline{\psi}_-, \\
             \left[Q_+, T_{++++} \right] &= \frac{i}{2} \partial_{++} S_{+++}, \\
              \left[Q_+, T_{++--} \right] &= \frac{i}{2} \partial_{++} S_{+--}, \\
              \left[Q_+, T_{----} \right] &= - \frac{i}{2}\partial_{--} S_{+--}, \\
               \left[Q_+, Y_{++} \right] &= i \sqrt{2} \partial_{++} \psi_+,  \\ 
            \left[Q_+, Y_{--} \right] &= i \sqrt{2} \partial_{--} \psi_+, \\ 
            \left[Q_+, \overline{Y}_{++} \right] &= 0, \\ 
            \left[Q_+, \overline{Y}_{--} \right] &= 0, \\
            \left[Q_+, G_{++} \right] &= \partial_{++} S_{-++} - i \sqrt{2} \partial_{++} \overline{\psi}_+,  \\
            \left[Q_+, G_{--} \right] &= \partial_{--} S_{-++} - i \sqrt{2} \partial_{--} \overline{\psi}_+, \\
            \left[Q_+, \overline{G}_{++} \right] &= 0, \\
            \left[Q_+. \overline{G}_{--} \right] &= 0, \\ 
                  \left\{Q_+, S_{+++} \right\} &= 0, \\
        \left\{Q_+, S_{+--} \right\} &= 0, \\
        \left\{Q_+, S_{-++} \right\} &= - i \overline{Y}_{++}, \\
        \left\{Q_+, S_{---} \right\} &= i \overline{Y}_{--}, \\
        \left\{Q_+, \oS_{+++} \right\} &= \frac{i}{2} \partial_{++} j_{++} - T_{++++}, \\
        \left\{Q_+, \oS_{+--} \right\} &= - T_{++--} - \frac{i}{2} \partial_{--} j_{++}, \\
        \left\{Q_+, \oS_{-++} \right\} &= i \overline{G}_{++}, \\
        \left\{Q_+, \oS_{---} \right\} &= -i \overline{G}_{--}.
    \end{aligned}
\end{equation}

\newpage 

For $Q_-$:
\begin{equation}
    \begin{aligned}
        \left[ Q_-, j_{++} \right] &= S_{-++} - i 2 \sqrt{2} \overline{\psi}_+, \\
        \left[Q_-, j_{--}\right] &= S_{---}, \\
        \left[Q_-, T_{++++}\right] &= - \frac{i}{2} \partial_{++} S_{-++}, \\
        \left[Q_-, T_{++--} \right] &= \frac{i}{2} \partial_{--} S_{-++} ,\\
        \left[Q_-, T_{----} \right] &= \frac{i}{2} \partial_{--} S_{---}, \\
        \left[Q_-, Y_{++} \right] &= i \sqrt{2} \partial_{++} \psi_-, \\
        \left[Q_-, Y_{--} \right] &= i \sqrt{2} \partial_{--} \psi_-, \\
        \left[Q_-, \overline{Y}_{++} \right] &= 0, \\
        \left[Q_-, \overline{Y}_{--} \right] &= 0, \\ 
          \left[Q_-, G_{++} \right] &= 0, \\
           \left[Q_-, G_{--} \right] &= 0, \\
            \left[Q_-, \overline{G}_{++} \right] &= \partial_{++} S_{+--} + i \sqrt{2} \partial_{++} \overline{\psi}_-, \\
        \left[Q_-, \overline{G}_{--} \right] &= \partial_{--} S_{+--} + i \sqrt{2} \partial_{--} \overline{\psi}_-, \\
        \left\{ Q_-, S_{+++}\right\} &= - i \overline{Y}_{++}, \\
        \left\{Q_-, S_{+--} \right\} &= i \overline{Y}_{--},\\
        \left\{Q_-, S_{-++}\right\} &= 0, \\
        \left\{Q_-, S_{---} \right\} &= 0, \\
         \left\{Q_-, \overline{S}_{+++} \right\} &= - i G_{++}, \\
        \left\{Q_-, \overline{S}_{+--} \right\} &= i G_{--}, \\
         \left\{Q_-, \overline{S}_{-++} \right\} &= - T_{++--} - \frac{i}{2} \partial_{++} j_{--}, \\  \left\{Q_-, \overline{S}_{---} \right\} &= - T_{----} + \frac{i}{2} \partial_{--} j_{--}.
    \end{aligned}
\end{equation}

\newpage 

For $\overline{Q}_+$:
\begin{equation}
    \begin{aligned}
        \left[\overline{Q}_+, j_{++} \right] &= - \overline{S}_{+++}, \\ 
        \left[\overline{Q}_+, j_{--} \right] &= - \overline{S}_{+--} + i 2\sqrt{2} \psi_-, \\
        \left[\overline{Q}_+, T_{++++} \right] &= \frac{i}{2} \partial_{++} \overline{S}_{+++}, \\
        \left[\overline{Q}_+, T_{++--} \right] &= \frac{i}{2} \partial_{++} \overline{S}_{+--}, \\
        \left[\overline{Q}_+, T_{----} \right] &= - \frac{i}{2} \partial_{--} \overline{S}_{+--}, \\
        \left[\overline{Q}_+, Y_{++} \right] &= 0, \\
        \left[\overline{Q}_+, Y_{--} \right] &= 0, \\
        \left[\overline{Q}_+, \overline{Y}_{++} \right] &= i \sqrt{2} \partial_{++} \overline{\psi}_+, \\
        \left[ \overline{Q}_+, \overline{Y}_{--}\right] &= i \sqrt{2} \partial_{--} \overline{\psi}_+, \\
        \left[\overline{Q}_+, G_{++} \right] &= 0, \\
        \left[ \overline{Q}_+, G_{--} \right] &= 0, \\
        \left[ \overline{Q}_+, \overline{G}_{++} \right] &= - \partial_{++} \overline{S}_{-++} - i \sqrt{2} \partial_{++} \psi_+, \\
        \left[\overline{Q}_+, \overline{G}_{--}\right] &=  - \partial_{--} \overline{S}_{-++} - i \sqrt{2}\partial_{--} \psi_+, \\
              \left\{\overline{Q}_+, S_{+++} \right\} &= - T_{++++} - \frac{i}{2} \partial_{++} j_{++},
        \\ \left\{\overline{Q}_+, S_{+--} \right\} &= - T_{++--} + \frac{i}{2}\partial_{--} j_{++}, \\
        \left\{\overline{Q}_+, S_{-++} \right\} &= - i G_{++}, \\
        \left\{\overline{Q}_+, S_{---} \right\} &= i G_{--}, \\
        \left\{\overline{Q}_+, \overline{S}_{+++} \right\} &= 0, \\
        \left\{\overline{Q}_+, \overline{S}_{-++} \right\} &= i Y_{++}, \\
        \left\{\overline{Q}_+, \overline{S}_{---} \right\} &= - i Y_{--}.
    \end{aligned}
\end{equation}

\newpage

For $\overline{Q}_-$:
\begin{equation}
    \begin{aligned}
        \left[\overline{Q}_-, j_{++} \right] &= - \overline{S}_{-++} - i 2\sqrt{2} \psi_+, \\
        \left[\overline{Q}_-, j_{--} \right] &= - \overline{S}_{---}, \\
        \left[\overline{Q}_-, T_{++++} \right] &= - \frac{i}{2} \partial_{++} \overline{S}_{-++}, \\
        \left[\overline{Q}_-, T_{++--} \right] &= \frac{i}{2} \partial_{--} \overline{S}_{-++}, \\
        \left[\overline{Q}_-, T_{----} \right] &= \frac{i}{2} \partial_{--} \overline{S}_{---}, \\
        \left[\overline{Q}_-, Y_{++} \right] &= 0, \\
        \left[\overline{Q}_-, Y_{--} \right] &= 0, \\
        \left[\overline{Q}_-, \overline{Y}_{++} \right] &= i \sqrt{2} \partial_{++} \overline{\psi}_-, \\
        \left[\overline{Q}_-, \overline{Y}_{--} \right] &= i \sqrt{2} \partial_{--} \overline{\psi}_-, \\
        \left[\overline{Q}_-, G_{++} \right] &= - \partial_{++} \overline{S}_{+--} + i \sqrt{2} \partial_{++} \psi_-,  \\
        \left[\overline{Q}_-, G_{--} \right] &= - \partial_{--} \overline{S}_{+--} + i \sqrt{2} \partial_{--} \psi_-,  \\
        \left[\overline{Q}_-, \overline{G}_{++} \right] &= 0, \\
        \left[Q_-, \overline{G}_{--} \right] &= 0,\\
        \left\{\overline{Q}_-, S_{+++} \right\} &= i \overline{G}_{++}, \\ 
        \left\{\overline{Q}_-, S_{+--} \right\} &= - i \overline{G}_{--}, \\ 
        \left \{\overline{Q}_-, S_{-++} \right\} &= - T_{++--} + \frac{i}{2} \partial_{++} j_{--}, \\
        \left \{\overline{Q}_-, S_{---} \right\} &= - T_{----} - \frac{i}{2} \partial_{--} j_{--}, \\
        \left\{\overline{Q}_-, \overline{S}_{+++} \right\} &= i Y_{++}, \\
        \left\{\overline{Q}_-, \overline{S}_{+--} \right\} &= - i Y_{--}, \\
        \left\{\overline{Q}_-, \overline{S}_{-++} \right\} &= 0, \\
        \left \{ \overline{Q}_-, \overline{S}_{---}\right \} &= 0.
    \end{aligned}
\end{equation}

\section{\texorpdfstring{$T\oT$}{} deformed \texorpdfstring{$\mathcal{N} = (2,2)$ $\mS$}{}-multiplet}
\label{App:22}
In this appendix, we collect our deformed $\mathcal{N} = (2, 2)$ $\mS$-multiplet results.

The elements of the deformed $\mathcal{N} = (2, 2)$ $\mS$-multiplet which vanishes at the superconformal point are:

\begin{equation}
\begin{aligned}
	\psi_+ &= - \frac{i\pi\lambda}{32\sqrt{2}} \bigg(\frac{i}{2}\partial_{++} j_{++}^{(0)} - T_{++++}^{(0)}\bigg) \oS_{---}^{(0)} + O(\lambda^2), \\
    \psi_- &= + \frac{i\pi\lambda}{32\sqrt{2}} \bigg(\frac{i}{2}\partial_{--} j_{--}^{(0)} - T_{----}^{(0)}\bigg) \oS_{+++}^{(0)} + O(\lambda^2), \\
    \opsi_+ &= - \frac{i\pi\lambda}{32\sqrt{2}} \bigg(\frac{i}{2}\partial_{++} j_{++}^{(0)} + T_{++++}^{(0)} \bigg)  S_{---}^{(0)} + O(\lambda^2), \\
    \opsi_- &= + \frac{i\pi\lambda}{32\sqrt{2}}\bigg(\frac{i}{2}\partial_{--} j_{--}^{(0)} + T_{----}^{(0)} \bigg) S_{+++}^{(0)} + O(\lambda^2),
    \end{aligned}
    \end{equation}
    
    \begin{equation}
        \begin{aligned}
        S_{+--} &= \frac{\pi\lambda}{16} S_{+++}^{(0)} T_{----}^{(0)} + O(\lambda^2), \\
    S_{-++} &= \frac{\pi\lambda}{16} S_{---}^{(0)} T_{++++}^{(0)} + O(\lambda^2), \\
    \oS_{+--} &= \frac{\pi\lambda}{16} \oS_{+++}^{(0)} T_{----}^{(0)} + O(\lambda^2), \\
    \oS_{-++} &= \frac{\pi\lambda}{16} \oS_{---}^{(0)} T_{++++}^{(0)} + O(\lambda^2), 
    \end{aligned}
    \end{equation}
    
    \begin{equation}
        \begin{aligned}
    Y_{++} &= \frac{\pi\lambda}{32} \partial_{++} \oS_{+++}^{(0)} \oS_{---}^{(0)} + O(\lambda^2), \\
    Y_{--} &= \frac{\pi\lambda}{32} \oS_{+++}^{(0)} \partial_{--} \oS_{---}^{(0)} + O(\lambda^2), \\
    \oY_{++} &= - \frac{\pi\lambda}{32} \partial_{++} S_{+++}^{(0)} S_{---}^{(0)} + O(\lambda^2), \\
    \oY_{--} &= - \frac{\pi\lambda}{32} S_{+++}^{(0)} \partial_{--} S_{---}^{(0)} + O(\lambda^2),
    \end{aligned}
    \end{equation}
  
    \begin{equation}
    \begin{aligned}
    G_{++} &= - \frac{\pi\lambda}{32} \partial_{++} \oS_{+++}^{(0)} S_{---}^{(0)} + O(\lambda^2), \\
    G_{--} &= - \frac{\pi\lambda}{32} \oS_{+++}^{(0)} \partial_{--} S_{---}^{(0)} + O(\lambda^2), \\
    \oG_{++} &= \frac{\pi\lambda}{32} \partial_{++} S_{+++}^{(0)} \oS_{---}^{(0)} + O(\lambda^2), \\
    \oG_{--} &= \frac{\pi\lambda}{32} S_{+++}^{(0)} \partial_{--} \oS_{---}^{(0)} + O(\lambda^2). 
\end{aligned}
\end{equation}

The leading order correction of the operators which do not vanish at the superconformal point can be solved from
\begin{equation}
\begin{aligned}
	\partial_{++} j_{--} &= - \frac{\pi\lambda}{16} \partial_{--} j^{(0)}_{--} T_{++++}^{(0)} + O(\lambda^2), \\
    \partial_{--} j_{++} &= - \frac{\pi\lambda}{16} \partial_{++} j^{(0)}_{++} T_{----}^{(0)} + O(\lambda^2), \\
    \partial_{--} S_{+++} &= - \frac{\pi\lambda}{16} \partial_{++} S_{+++}^{(0)} T_{----}^{(0)} + O(\lambda^2), \\
    \partial_{++} S_{---} &= - \frac{\pi\lambda}{16} \partial_{--} S_{---}^{(0)} T_{++++}^{(0)} + O(\lambda^2), \\
    \partial_{--} \oS_{+++} &= - \frac{\pi\lambda}{16} \partial_{++} \oS_{+++}^{(0)} T_{----}^{(0)} + O(\lambda^2), \\
    \partial_{++} \oS_{---} &= - \frac{\pi\lambda}{16} \partial_{--} \oS_{---}^{(0)} T_{++++}^{(0)} + O(\lambda^2), \\
    \partial_{--} T_{++++} &= - \frac{\pi\lambda}{16} \partial_{++}T_{++++}^{(0)} T_{----}^{(0)} + O(\lambda^2), \\
    \partial_{++} T_{----} &= - \frac{\pi\lambda}{16} \partial_{--} T_{----}^{(0)} T_{++++}^{(0)} + O(\lambda^2). 
\end{aligned}
\end{equation}

At first glance, the deformation breaks both $U(1)_V$ and $U(1)_A$ $R$-symmetries, however, any one of the symmetries can be restored by some improvement transformation, but we can restore only one of them. For instance, we can shift away the superfield $\mathcal{Y}$ to restore the $U(1)_V$ symmetry to improve the $\mathcal{S}$-multiplet into $\mathcal{R}$-multiplet. Consider the following improvement transformation,
\begin{equation}
\begin{aligned}
	\delta j_{++}   &= \frac{\pi\lambda}{16} j_{--}^{(0)} T_{++++}^{(0)}, \\
    \delta j_{--}   &= \frac{\pi\lambda}{16} j_{++}^{(0)} T_{----}^{(0)}, \\
    \delta S_{+++}  &=\frac{i\pi\lambda}{32} j_{--}^{(0)} \partial_{++} S_{+++}^{(0)}, \\
    \delta S_{+--}  &= -\frac{i\pi\lambda}{32} \partial_{--} j_{--}^{(0)} S_{+++}^{(0)}, \\
    \delta\oS_{+++} &=-\frac{i\pi\lambda}{32} j_{--}^{(0)} \partial_{++}\oS_{+++}^{(0)},\\
    \delta\oS_{+--} &= \frac{i\pi\lambda}{32} \partial_{--} j_{--}^{(0)}\oS_{+++}^{(0)},\\
    \delta S_{-++}  &=-\frac{i\pi\lambda}{32} \partial_{++} j_{++}^{(0)} S_{---}^{(0)}, \\
    \delta S_{---}  &= \frac{i\pi\lambda}{32} j_{++}^{(0)} \partial_{--} S_{---}^{(0)}, \\
    \delta \oS_{-++}&= \frac{i\pi\lambda}{32} \partial_{++} j_{++}^{(0)}\oS_{---}^{(0)},\\
    \delta \oS_{---}&=-\frac{i\pi\lambda}{32} j_{++}^{(0)} \partial_{--}\oS_{---}^{(0)},\\
    \delta Y_{++}   &= \frac{\pi\lambda}{32} \oS_{---}^{(0)} \partial_{++} \oS_{+++}^{(0)}, \\
    \delta Y_{--}   &=-\frac{\pi\lambda}{32} \oS_{+++}^{(0)} \partial_{--} \oS_{---}^{(0)}, \\
    \delta \oY_{++} &=-\frac{\pi\lambda}{32} S_{---}^{(0)} \partial_{++} S_{+++}^{(0)}, \\
    \delta \oY_{--} &= \frac{\pi\lambda}{32} S_{+++}^{(0)} \partial_{--} S_{---}^{(0)}, \\
    \delta G_{++}   &= \frac{\pi\lambda}{32} S_{---}^{(0)} \partial_{++} \oS_{+++}^{(0)}, \\
    \delta G_{--}   &=-\frac{\pi\lambda}{32} \oS_{+++}^{(0)} \partial_{--} S_{---}^{(0)}, \\
    \delta \oG_{++} &=-\frac{\pi\lambda}{32} \oS_{---}^{(0)} \partial_{++} S_{+++}^{(0)}, \\
    \delta \oG_{--} &= \frac{\pi\lambda}{32} S_{+++}^{(0)} \partial_{--} \oS_{---}^{(0)}, \\
    \delta \psi_+   &= \frac{i\pi\lambda}{32\sqrt{2}}\bigg(\frac{i}{2}\partial_{++} j_{++}^{(0)} - T_{++++}^{(0)}\bigg) \oS_{---}^{(0)}, \\
    \delta \psi_-   &=-\frac{i\pi\lambda}{32\sqrt{2}}\bigg( \frac{i}{2}\partial_{--}j^{(0)}_{--}-T^{(0)}_{----}\bigg)\oS_{+++}^{(0)}, \\
    \delta \opsi_+    &= \frac{i\pi\lambda}{32\sqrt{2}}\bigg(T_{++++}^{(0)} + \frac{i}{2}\partial_{++} j_{++}^{(0)}\bigg)S_{---}^{(0)}, \\
    \delta \opsi_-    &=-\frac{i\pi\lambda}{32\sqrt{2}}\bigg(T_{++++}^{(0)} + \frac{i}{2}\partial_{++}j_{++}^{(0)}\bigg) S_{+++}^{(0)},
\end{aligned}
\end{equation}
\begin{equation}
\begin{aligned}
    \delta T_{++++} &= -\frac{\pi\lambda}{64} j_{--}^{(0)} \partial_{++}^2 j_{++}^{(0)}, \\
    \delta T_{++--} &= +\frac{\pi\lambda}{64}\partial_{--} j_{--}^{(0)} \partial_{++} j_{++}^{(0)}, \\
    \delta T_{----} &= -\frac{\pi\lambda}{64} j_{++}^{(0)} \partial_{--}^2 j_{--}^{(0)}.
\end{aligned}
\end{equation}

After the improvement transformation, we find
\begin{equation}
	\psi_\pm = \opsi_\pm = Y_{\pm\pm} = \oY_{\pm\pm} = 0
\end{equation}
and 
\begin{equation}
\begin{aligned}
	S_{+--} &= \frac{\pi\lambda}{16}\bigg(T_{----}^{(0)} - \frac{i}{2}\partial_{--}j_{--}\bigg)S_{+++}^{(0)} + O(\lambda^2), \\
    \oS_{+--} &=\frac{\pi\lambda}{16}\bigg(T_{----}^{(0)} + \frac{i}{2}\partial_{--}j_{--}\bigg)\oS_{+++}^{(0)} + O(\lambda^2), \\
    S_{-++} &= \frac{\pi\lambda}{16}\bigg(T_{++++}^{(0)} - \frac{i}{2}\partial_{++}j^{(0)}_{++}\bigg)S_{---}^{(0)} + O(\lambda^2), \\
    \oS_{-++} &= \frac{\pi\lambda}{16}\bigg(T_{++++}^{(0)} + \frac{i}{2}\partial_{++}j_{++}^{(0)}\bigg)\oS_{---}^{(0)} + O(\lambda^2), \\
    T_{++--} &= \frac{\pi\lambda}{16} T_{++++}^{(0)} T_{----}^{(0)} + \frac{\pi\lambda}{64}\partial_{--}j_{--}^{(0)} \partial_{++}j_{++}^{(0)} + O(\lambda^2), \\
    G_{++} &= -\frac{\pi\lambda}{16} \partial_{++} \oS_{+++}^{(0)}S_{---}^{(0)} + O(\lambda^2), \\
    G_{--} &= -\frac{\pi\lambda}{16} \oS_{+++}^{(0)} \partial_{--} S_{---}^{(0)} + O(\lambda^2), \\
    \oG_{++} &= \frac{\pi\lambda}{16} \partial_{++} S_{+++}^{(0)} \oS_{---}^{(0)} + O(\lambda^2), \\
    \oG_{--} &= \frac{\pi\lambda}{16} S_{+++}^{(0)} \partial_{--} \oS_{---}^{(0)} + O(\lambda^2).
\end{aligned}
\end{equation}

\bibliographystyle{JHEP}
\bibliography{ref}
\end{document}